\documentclass{aa}
\usepackage{psfig}
\newfont{\sfsl}{cmssqi8 scaled 1200}

\begin{document}

\title{A New Measurement of the X-ray Temperature Function of Clusters of Galaxies}

\subtitle{}

\author{Yasushi Ikebe\inst{1}, 
	Thomas H. Reiprich\inst{2}, 
	Hans B\"{o}hringer\inst{1},
	Yasuo Tanaka\inst{1,3}
	\and
	Tetsu Kitayama\inst{4}
	}

\offprints{Y. Ikebe}
\mail{ikebe@mpe.mpg.de}

\institute{Max-Planck-Institut f\"{u}r extraterrestrische Physik,
           Postfach 1312, 85741 Garching, Germany
	   \and
	   Department of Astronomy, University of Virginia,
	   PO Box 3818, 530 McCormick Road, Charlottesville, 
	   VA 22903-0818, USA
	   \and
           Institute of Space and Astronautical Science,
           Yoshinodai 3-1-1, Sagamihara, Kanagawa 229-8510, Japan
	   \and
	   Department of Physics, Toho University,
  	   Miyama, Funabashi, Chiba 274-8510, Japan
           }

\date{Received month day, 2001; accepted month day, 2001}

\authorrunning{Ikebe et al.}

\label{firstpage}

\abstract{
We present a newly measured X-ray temperature function of galaxy clusters
using a complete flux-limited sample of 61 clusters.
The sample is constructed with the total survey area of 8.14 steradians
and the flux limit of $1.99\times 10^{-11}$ ergs s$^{-1}$ cm$^{-2}$
in the 0.1--2.4~keV band.
X-ray temperatures and fluxes of the sample clusters
were accurately measured with {\it ASCA} and {\it ROSAT} data.
The derived temperature function covers an unprecedentedly wide temperature
range of 1.4-11~keV.
By fitting these data with theoretically predicted temperature functions
given by the Press-Schechter formalism together with a recent formation
approximation and the CDM power spectrum,
we obtained tight and individual constraints on 
$\Omega_{\rm m,0}$ and $\sigma_8$.
We also employed the Formation-Epoch model
in which the distribution in the formation epoch of clusters
as well as the temperature evolution are taken into account,
showing significantly different results.
Systematics caused by the uncertainty in the mass-temperature relation are studied
and found to be as large as the statistical errors.
}

\maketitle

%====1====1====1====1====1====1====1====1====1====1====1====
\section{Introduction}
%====1====1====1====1====1====1====1====1====1====1====1====
%-- Cosmology with Mass Function of clusters --\\
The mass function of clusters of galaxies (MF),
the number density of the most massive virialized systems,
contains information on the structure formation history of the universe.
A theoretical framework, e.g. the Press-Schechter formalism
(Press \& Schechter \cite{PS}) together with the Cold Dark Matter model,
has been established to predict the MF.
This allows us to constrain cosmological parameters 
using an observationally determined MF for the present epoch
(as well as its time evolution for even tighter constraints).
In particular, $\sigma_8$, the amplitude of mass density fluctuations
on a scale of 8$h^{-1}$ Mpc where $h$ is the Hubble constant in
unit of 100 km/s/Mpc,
and $\Omega_{\rm m,0}$, the mean matter density,
are most sensitively determined by the cluster abundance measurements
(e.g. Henry \& Arnaud \cite{HA};
White et al. \cite{WEF};
Eke et al. \cite{ECF}; Kitayama \& Suto \cite{KS};
Viana \& Liddle \cite{VL}; Oukbir \& Blanchard \cite{OB};
Pen \cite{Pen}; Eke et al. \cite{Eke}).

%-- Temperature Function instead of Mass Function --\\
Observationally the local MF has been derived from
measuring masses of individual clusters
from galaxy velocity dispersions or other optical properties
by Bahcall and Cen (\cite{BC}), Biviano et al. (\cite{Biviano}), and
Girardi et al. (\cite{Girardi}).
The estimated virial masses for individual clusters
depend rather strongly on model assumptions, however.
As argued by Evrard et al. (\cite{Evrard97}) on the basis of hydrodynamical
N-body simulations, cluster masses may be presently more accurately 
determined from a temperature measurement 
and a mass-temperature relation determined from detailed observations
or numerical modeling.
Thus alternatively, as a well-defined observational quantity,
the X-ray temperature function (XTF) has been measured,
which can be converted to the MF by means of the mass-temperature relation.

%-- Short history of XTF measuremnt and its improvement --\\
The first measurements of the XTF were reported
by Edge et al. (\cite{Edge}) and Henry \& Arnaud (\cite{HA}),
using an X-ray flux-limited sample of 45 clusters and 25 clusters,
respectively.
Recent observational improvement on the XTF was made 
by Markevitch (\cite{Markevitch}), Henry (\cite{Henry}),
Blanchard et al. (\cite{Blanchard}), and Pierpaoli et al. (\cite{Pierpaoli}),
using more accurate temperature-measurement results
for each cluster with {\it ASCA} data (Tanaka et al. \cite{ASCA}).
However, the narrow temperature ranges (3-10~keV) in which
the XTF is defined so far do not allow 
an investigation of a more detailed shape of the XTF
other than just fitting a single power-law.
Without better information on the actual shape of the XTF,
no independent constraints on $\sigma_8$ and $\Omega_{\rm m,0}$
can be derived, and a wide range of combinations is still allowed.
Recently, a number of new X-ray-cluster surveys were performed,
which provide a high completeness for the brightest clusters.
This motivated us to revisit the measurement of the local XTF
and to improve its accuracy in order to derive narrower constraints
on the cosmological parameters.

Reiprich \& B\"{o}hringer (\cite{RB}) have compiled
a new X-ray flux-limited cluster sample ({\sfsl HIFLUGCS})
with a flux limit of $2\times10^{-11}$ ergs s$^{-1}$ cm$^{-2}$ (0.1-2.4~keV).
Compared with previous samples with similar flux limits,
their catalog covers the largest volume and is the most complete.
Based on this cluster sample, Reiprich \& B\"{o}hringer have measured
total masses for individual clusters and derived the X-ray mass function
for the first time.
In this paper we, using this cluster sample,
report the construction of a new measurement of the XTF.
The larger number of clusters in the sample
together with accurate temperature measurements with {\it ASCA} data
leads to a significant improvement of the XTF,
which covers 1.4-11~keV temperature range.
The X-ray flux-limited sample and the new temperature measurements
with {\it ASCA} data are presented in Sect. 2.
The derivation of the XTF is described in Sect. 3.
In Sect. 4, the XTF is used to obtain constraints on cosmological parameters, 
and the results are compared with other work in Sect. 5.
Throughout the paper, the Hubble constant is given as $100\ h$ km/s/Mpc,
and $\log$ and $\ln$ denotes a decimal and natural logarithm, respectively.

%====2=====2=====2=====2=====2=====2=====2=====2=====2=====
\section{Sample and analysis of {\it ASCA} data}
%====2=====2=====2=====2=====2=====2=====2=====2=====2=====
%----2.1---2.1---2.1---2.1---2.1---2.1---2.1---2.1---2.1----
\subsection{Master catalog}
%----2.1---2.1---2.1---2.1---2.1---2.1---2.1---2.1---2.1----
%-- 1.7x10**-11 sample, PSPC count rates --\\
For this study we use a sample of 106 galaxy clusters compiled by
Reiprich \& B\"{o}hringer (\cite{RB}),
which is used to construct a flux-limited complete sample
({\sfsl HIFLUGCS}) with the flux-limit
of $2.0\times10^{-11}$ ergs s$^{-1}$ cm$^{-2}$ (0.1-2.4~keV).
The 106 clusters have been derived from combining previous X-ray cluster
or elliptical galaxy catalogs 
that include B\"{o}hringer et al. (\cite{NORASI}, \cite{REFLEXI}), 
B\"{o}hringer (\cite{hxb99}),
Retzlaff et al. in preparation,
Ebeling et al. (\cite{XBACs}, \cite{BCS}), de Grandi et al. (\cite{deGrandi}),
Beuing et al. (\cite{Beuing}), Lahav et al. (\cite{Lahav}), 
and Edge et al. (\cite{Edge}).
Most of them are based on the {\it ROSAT} All Sky Survey
(Tr\"{u}mper \cite{rosat}; Voges et al. \cite{rass}).
From these catalogs, all objects satisfying certain criteria,
mainly that the measured flux given in any of these catalogs
is above $1.7\times10^{-11}$ ergs s$^{-1}$ cm$^{-2}$ (0.1-2.4~keV),
have been sampled.
A defined region around the center of M87 is not included 
in the survey area,
because of the large scale emission of the Virgo cluster
that compromises cluster detection and characterization in the region.
This excludes the Virgo cluster and M86 from the master catalog.
The count rate of the PSPC (Pfeffermann et al. \cite{PSPC})
in the 0.1--2.4~keV band for each object was redetermined
from the {\it ROSAT} All Sky Survey data or PSPC pointing observations
whenever available
by integrating all X-ray flux within a maximum radius where
the cluster emission is significantly detected.
See Reiprich \& B\"{o}hringer (\cite{RB}) for detailed descriptions
of the sample construction and data analysis.
The names and the redshifts compiled from the recent literature
of all the 106 clusters are listed in Table 1.

%-- construct 1.99x10**-11 sample --\\
As described below,
we newly determine the X-ray temperature as well as the X-ray flux
for each of the 106 clusters.
{\it ASCA} data are available for 88 of the 106 clusters,
and the X-ray temperatures are measured by analyzing the {\it ASCA} data
(Sect. 2.2).
Using the {\it ASCA} results,
the correlation between the X-ray luminosity and the temperature,
$L-T$ relation, is established (Sect. 2.3).
Among the other clusters for which no {\it ASCA} data exist
(18 clusters), temperatures of 4 clusters have been measured with
{\it Einstein}, {\it EXOSAT}, or {\it XMM-Newton} in previous works
and were used in our analysis. For the remaining 14 clusters,
the temperatures are estimated from the observed {\it ROSAT} PSPC
count rate by means of the $L-T$ relation.
By setting a flux-limit of $1.99\times10^{-11}$ ergs s$^{-1}$ cm$^{-2}$ (0.1--2.4~keV),
a complete flux-limited sample comprising 63 clusters
is then constructed (Sect. 2.4).

%----2.2---2.2---2.2---2.2---2.2---2.2---2.2---2.2---2.2----
\subsection{Temperature Measurements}
%----2.2---2.2---2.2---2.2---2.2---2.2---2.2---2.2---2.2----
%-- Central cool component has to be separated --\\
For our study we need to determine an average temperature
for each cluster that closely reflects the mass of the cluster,
i.e. a virial temperature.
In many clusters, the X-ray emitting hot gas is not isothermal
but a cooler gas component is often observed in the central region.
The cool component, which is due to either a cooling flow
(e.g. Fabian \cite{CoolingFlow}) or the ISM of cD galaxies 
(e.g. Makishima et al. \cite{max01}),
often exhibits a significant fraction of the luminosity
and has to be separated from the rest of the X-ray emission
to determine the cluster average temperature
(see also Markevitch \cite{Markevitch}; Arnaud \& Evrard \cite{AE}).

%-- We used two temperature model --\\
Therefore, in the {\it ASCA} data analysis,
we employed a two-temperature (2T) model,
in which isothermal plasma is filling the entire cluster region,
while, in the central region, another cooler isothermal gas component
is allowed to coexist with the hotter plasma forming a multi-phase 
intra-cluster medium (ICM).
This 2T picture was established for the Centaurus cluster
(Fukazawa et al. \cite{FukazawaCen}; Ikebe et al. \cite{IkebeCen})
and also gave a good account of the {\it ASCA} data of Virgo/M87 
(Matsumoto et al. \cite{MatsumotoM87}),
Hydra-A (Ikebe et al. \cite{IkebeHydra}),
Abell~1795 (Xu et al. \cite{XuA1795}),
and other nearby clusters (Fukazawa \cite{FukazawaPhD};
Fukazawa et al. \cite{Fukazawa98}, \cite{Fukazawa00}).
The hot component extends over the major part of cluster volume
as far as is measurable
(e.g. Fukazawa \cite{FukazawaPhD}; White \cite{White00}).
The temperature of this hot component derived from the 2T model
therefore represents the best estimation of the virial temperature 
of the cluster.
Although a cooling flow spectral model also gives generally 
a good description of the {\it ASCA} spectra
(e.g. Fabian et al. \cite{FABT94}; 
White \cite{White00}; Allen et al. \cite{Allen01}),
the estimated temperatures from where gas starts to cool
vary considerably from author to author
and are sometimes spuriously high.
More importantly, recent {\it XMM-Newton} data clearly show
that the conventional cooling flow spectral model is not adequate 
(Peterson et al. \cite{Peterson01}; Tamura et al. \cite{Tamura01}; 
Kaastra et al. \cite{Kaastra01}).
The 2T model, on the other hand, is valid to the {\it XMM-Newton} data
and the hot component temperature well represents the temperature of
the outer main component (see Ikebe \cite{Ikebe01}).

%-- non-isothermality besides central cool component --\\
It is known that, apart from the central cool component,
many clusters show deviations from isothermality 
such as an asymmetric temperature 
distribution due to merging (e.g. Briel \& Henry \cite{BH94})
or a global temperature decrement towards the outside
(Markevitch et al. \cite{MarkevitchTmap}).
For such clusters, the 2T model fit
gives insignificant flux to the cool component
and works practically as an isothermal model.
Thus the hot component temperature gives the average temperature,
which also can be a good measure of the virial temperature.

%-- collecting ASCA archive --\\
Among the 106 sample clusters,
88 clusters have been observed with {\it ASCA},
and all the data sets have become publicly available by now.
We retrieved the {\it ASCA} data 
from the {\it ASCA} archival data base provided
by NASA Goddard Space Flight Center and by Leicester University.
Here we will give a brief explanation of the analysis method 
of the {\it ASCA} data.

%-- Detailed ASCA data analysis procedure --\\
{\it ASCA} has four focal plane instruments,
two SIS (Solid-state Imaging Spectrometer)
and two GIS (Gas Imaging Spectrometer; Ohashi et al. \cite{OhashiGIS}),
which were usually used simultaneously to observe an astronomical object.
In all the observations, the GIS were operated in the normal PH mode
and the SIS were operated in FAINT mode or BRIGHT mode.
The number of active CCD chips of each SIS used was 4, 2, or 1,
corresponding to the field of view of $22'\times22'$, $22'\times11'$,
or $11'\times11'$, respectively.

We discarded the GIS and SIS data that were taken when
the elevation angle of the X-Ray Telescope (XRT) from the local horizon
was less than $5^{\circ}$.
An additional screening requirement, that the elevation angle from
the sunlit earth be greater than $25^{\circ}$ and $20^{\circ}$,
was applied to the SIS-0 and SIS-1 data, respectively.
In order to ensure a low and stable particle background,
we also discarded GIS and SIS data acquired under a geomagnetic 
cutoff rigidity smaller than 6 GV.

The background is composed of cosmic X-ray background
and non-X-ray background, which were estimated for each cluster observation
from the data of blank-sky observations as well as
the data taken when the X-Ray Telescope was pointing at the dark
(night) earth.

In order to maximize the detection efficiency of the central cool component,
we accumulated spectra over two regions, a central region of $2'$ radius 
and an outer region from $2'$ to $r_{\rm max}$, from each cluster data set.
$r_{\rm max}$, the maximum radius, is defined as the radius
within which at least 95\% of all detected source X-ray counts, 
within the field of view of the GIS, are included.
The data from the two GIS sensors and two SIS sensors were combined.
Thus for each cluster, four spectra: two central and two outer spectra
for GIS and SIS, were derived.

An effective area as a function of X-ray energy
was calculated for each spectrum with an {\it ASCA} simulator (SimARF).
The point-spread-function of the {\it ASCA} XRT is so largely extended
that the X-ray flux from the central brightest region may contribute
to the outer region spectrum
(Serlemitsos et al. \cite{XRT}; Takahashi et al. \cite{TakahashiASCANews}).
We therefore, simultaneously fitted the four spectra with the 2T model,
taking into account the flux-contamination effect.
As a plasma code, we used the MEKAL model
(Mewe et al. \cite{MGV85}, \cite{MLV86}; 
Kaastra \cite{Kaastra92}, Liedahl et al. \cite{LOG95}),
that is implemented in XSPEC version 10.0 (Arnaud \cite{XSPEC}).

A best-fit model was derived by a chi-square minimization method.
The cool and hot component temperatures are separate free parameters.
Many clusters do not require an additional cool component to fit
the spectra, and the cool component temperatures cannot be constrained.
For each cluster we investigated the significance of adding 
a cool component by comparison with a fit of a hot isothermal model
and performing an F-test.
For a cluster showing a low significance for the additional cool component
in the 2T model,
we fixed the temperature of the cool component at $T_{\rm iso}/2$, where
$T_{\rm iso}$ is the temperature derived with the isothermal model fitting.
The choice of the fixed cool-component temperature is based on
a clear correlation found in the 2T model fitting
that, for clusters showing a high significance for the cool component,
the cool-component temperature is always very close to half of 
the hot component temperature (see Ikebe \cite{Ikebe01}).
The hot component temperatures thus derived are summarized in Table 1.
A more detailed analysis procedure description and
results on the cool components will be presented in a following publication.

\begin{figure}
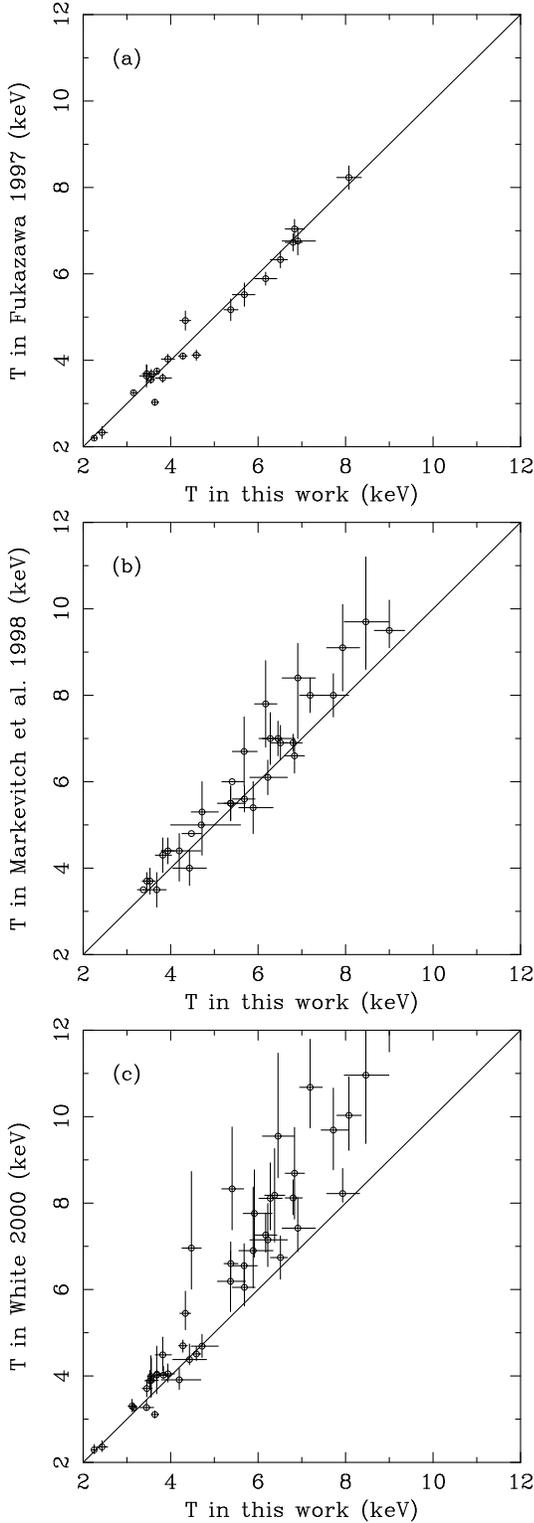

\centerline{
\psfig{file=h2920f1a.ps,width=7.0cm,angle=-90,clip=}}
\centerline{
\psfig{file=h2920f1b.ps,width=7.0cm,angle=-90,clip=}}
\centerline{
\psfig{file=h2920f1c.ps,width=7.0cm,angle=-90,clip=}}
\caption{Comparison between the cluster temperatures derived here
and those by Fukazawa (\cite{FukazawaPhD}) (a),
by Markevitch et al. (\cite{MarkevitchTmap}) (b), 
and by White (\cite{White00}) (c).}
\label{T-T}
\end{figure}

We compared the temperatures derived here with those of previous workers.
The temperatures agree with results of Fukazawa (\cite{FukazawaPhD})
within 1.5\% on average (Fig.~\ref{T-T}a),
who performed a similar analysis
on a smaller number of clusters with {\it ASCA} data,
accumulating spectra from a central region of 2-3 arc-minutes radius
and an outer region, and fitting them with the 2T model.
On the other hand, 
temperatures given by Markevitch et al. (\cite{MarkevitchTmap})
and White (\cite{White00}),
who used different (cooling flow) models to fit the {\it ASCA} spectra,
are higher than our values respectively by a factor of 1.09 and 1.25 on average
for clusters hotter than 6~keV (Fig.~\ref{T-T}b,c).
This difference causes a significant difference in the resulting XTFs 
(see Sect. 3.2).

\begin{table*}
\begin{center}
\caption{Summary of the cluster sample}
\begin{tabular}{cccrccc}
\hline %------------------------------------------------
\hline %------------------------------------------------
 NAME & redshift & $n_{\rm H}$
 & $kT_{\rm h}\ \ \ \ $
 & Flux(0.1-2.4keV)
 & \multicolumn{2}{c}{$^{\dag}$$L_{\rm X}$ (0.1-2.4keV) ($h^{-2}$ ergs s$^{-1}$ cm$^{-2}$)}\\
 & & ($10^{21}$cm$^{-2}$) & (keV)\ \ \ \ & ($10^{-11}$ergs s$^{-1}$ cm$^{-2}$) & Open & Flat \\
\hline %------------------------------------------------
 $^{*}$PERSEUS & 0.0183 &  1.569 & $ 6.42^{+ 0.06}_{-0.06}$ & $ 114.33\pm 1.50$  & ($ 4.12\pm 0.05$)e+44 & ($ 4.19\pm 0.05$)e+44 \\
 $^{*}$OPHIUCHUS & 0.0280 &  2.014 & $10.25^{+ 0.30}_{-0.36}$ & $  36.25\pm 1.16$  & ($ 3.07\pm 0.10$)e+44 & ($ 3.14\pm 0.10$)e+44 \\
       COMA & 0.0232 &  0.089 & $ 8.07^{+ 0.29}_{-0.27}$ & $  35.13\pm 0.79$  & ($ 2.04\pm 0.05$)e+44 & ($ 2.08\pm 0.05$)e+44 \\
 $^{*}$A3627 & 0.0163 &  2.083 & $ 5.62^{+ 0.12}_{-0.11}$ & $  31.60\pm 1.57$  & ($ 9.04\pm 0.45$)e+43 & ($ 9.16\pm 0.45$)e+43 \\
      A3526 & 0.0103 &  0.825 & $ 3.69^{+ 0.05}_{-0.04}$ & $  26.20\pm 0.95$  & ($ 2.99\pm 0.11$)e+43 & ($ 3.02\pm 0.11$)e+43 \\
 $^{*}$AWM7 & 0.0172 &  0.921 & $ 3.70^{+ 0.08}_{-0.05}$ & $  16.60\pm 0.55$  & ($ 5.30\pm 0.18$)e+43 & ($ 5.37\pm 0.18$)e+43 \\
 $^{*}$A2319 & 0.0564 &  0.877 & $ 8.84^{+ 0.29}_{-0.24}$ & $  12.15\pm 0.21$  & ($ 4.20\pm 0.07$)e+44 & ($ 4.39\pm 0.07$)e+44 \\
      A3571 & 0.0397 &  0.393 & $ 6.80^{+ 0.21}_{-0.18}$ & $  12.00\pm 0.15$  & ($ 2.05\pm 0.03$)e+44 & ($ 2.12\pm 0.03$)e+44 \\
 $^{*}$TRIANGUL & 0.0510 &  1.229 & $ 9.06^{+ 0.33}_{-0.31}$ & $  11.26\pm 0.13$  & ($ 3.18\pm 0.04$)e+44 & ($ 3.31\pm 0.04$)e+44 \\
      A2199 & 0.0302 &  0.084 & $ 4.28^{+ 0.10}_{-0.10}$ & $  10.73\pm 0.32$  & ($ 1.06\pm 0.03$)e+44 & ($ 1.09\pm 0.03$)e+44 \\
 $^{*}$3C129 & 0.0223 &  6.789 & $ 5.57^{+ 0.16}_{-0.15}$ & $  10.65\pm 0.98$  & ($ 5.72\pm 0.53$)e+43 & ($ 5.82\pm 0.54$)e+43 \\
      A1060 & 0.0114 &  0.492 & $ 3.15^{+ 0.05}_{-0.05}$ & $  10.04\pm 0.54$  & ($ 1.41\pm 0.07$)e+43 & ($ 1.42\pm 0.08$)e+43 \\
     2A0335 & 0.0349 &  1.864 & $ 3.64^{+ 0.09}_{-0.08}$ & $   8.87\pm 0.11$  & ($ 1.17\pm 0.01$)e+44 & ($ 1.21\pm 0.01$)e+44 \\
      A0262 & 0.0161 &  0.552 & $ 2.25^{+ 0.06}_{-0.06}$ & $   8.78\pm 0.55$  & ($ 2.46\pm 0.15$)e+43 & ($ 2.49\pm 0.16$)e+43 \\
     FORNAX & 0.0046 &  0.145 & $ 1.56^{+ 0.05}_{-0.07}$ & $   8.68\pm 0.66$  & ($ 1.97\pm 0.15$)e+42 & ($ 1.98\pm 0.15$)e+42 \\
      A0496 & 0.0328 &  0.568 & $ 4.59^{+ 0.10}_{-0.10}$ & $   8.20\pm 0.10$  & ($ 9.56\pm 0.12$)e+43 & ($ 9.81\pm 0.12$)e+43 \\
      A0085 & 0.0556 &  0.358 & $ 6.51^{+ 0.16}_{-0.23}$ & $   7.37\pm 0.07$  & ($ 2.48\pm 0.03$)e+44 & ($ 2.59\pm 0.03$)e+44 \\
      A3667 & 0.0560 &  0.459 & $ 6.28^{+ 0.27}_{-0.26}$ & $   7.12\pm 0.08$  & ($ 2.44\pm 0.03$)e+44 & ($ 2.54\pm 0.03$)e+44 \\
      A2029 & 0.0767 &  0.307 & $ 7.93^{+ 0.39}_{-0.36}$ & $   6.88\pm 0.07$  & ($ 4.43\pm 0.04$)e+44 & ($ 4.70\pm 0.05$)e+44 \\
      A3558 & 0.0480 &  0.363 & $ 5.37^{+ 0.17}_{-0.15}$ & $   6.68\pm 0.05$  & ($ 1.68\pm 0.01$)e+44 & ($ 1.74\pm 0.01$)e+44 \\
 $^{*}$S636 & 0.0116 &  0.642 & $ 2.06^{+ 0.07}_{-0.06}$ & $   6.65\pm 0.54$  & ($ 9.65\pm 0.79$)e+42 & ($ 9.75\pm 0.79$)e+42 \\
      A2142 & 0.0899 &  0.405 & $ 8.46^{+ 0.53}_{-0.49}$ & $   6.21\pm 0.09$  & ($ 5.52\pm 0.08$)e+44 & ($ 5.90\pm 0.08$)e+44 \\
      A1795 & 0.0616 &  0.120 & $ 6.17^{+ 0.26}_{-0.25}$ & $   6.21\pm 0.03$  & ($ 2.58\pm 0.01$)e+44 & ($ 2.70\pm 0.02$)e+44 \\
 $^{*}$PKS0745 & 0.1028 &  4.349 & $ 6.37^{+ 0.21}_{-0.20}$ & $   6.13\pm 0.10$  & ($ 7.18\pm 0.11$)e+44 & ($ 7.74\pm 0.12$)e+44 \\
      A2256 & 0.0601 &  0.402 & $ 6.83^{+ 0.23}_{-0.21}$ & $   6.03\pm 0.14$  & ($ 2.38\pm 0.05$)e+44 & ($ 2.49\pm 0.06$)e+44 \\
      A1367 & 0.0216 &  0.255 & $ 3.55^{+ 0.08}_{-0.08}$ & $   5.98\pm 0.07$  & ($ 3.02\pm 0.04$)e+43 & ($ 3.07\pm 0.04$)e+43 \\
      A3266 & 0.0594 &  0.148 & $ 7.72^{+ 0.35}_{-0.28}$ & $   5.77\pm 0.06$  & ($ 2.22\pm 0.02$)e+44 & ($ 2.32\pm 0.03$)e+44 \\
      A4038 & 0.0283 &  0.155 & $ 3.22^{+ 0.10}_{-0.10}$ & $   5.61\pm 0.12$  & ($ 4.86\pm 0.10$)e+43 & ($ 4.97\pm 0.10$)e+43 \\
      A2147 & 0.0351 &  0.329 & $ 4.34^{+ 0.12}_{-0.13}$ & $   5.45\pm 0.29$  & ($ 7.28\pm 0.39$)e+43 & ($ 7.49\pm 0.40$)e+43 \\
      A0401 & 0.0748 &  1.019 & $ 7.19^{+ 0.28}_{-0.24}$ & $   5.26\pm 0.09$  & ($ 3.23\pm 0.06$)e+44 & ($ 3.41\pm 0.06$)e+44 \\
      N5044 & 0.0090 &  0.491 & $ 1.22^{+ 0.04}_{-0.04}$ & $   5.20\pm 0.04$  & ($ 4.54\pm 0.04$)e+42 & ($ 4.58\pm 0.04$)e+42 \\
      A0478 & 0.0900 &  1.527 & $ 6.91^{+ 0.40}_{-0.36}$ & $   5.12\pm 0.05$  & ($ 4.58\pm 0.04$)e+44 & ($ 4.89\pm 0.05$)e+44 \\
    HYDRA-A & 0.0538 &  0.486 & $ 3.82^{+ 0.20}_{-0.17}$ & $   4.71\pm 0.04$  & ($ 1.49\pm 0.01$)e+44 & ($ 1.56\pm 0.01$)e+44 \\
      A2052 & 0.0348 &  0.290 & $ 3.12^{+ 0.10}_{-0.09}$ & $   4.57\pm 0.08$  & ($ 6.01\pm 0.10$)e+43 & ($ 6.18\pm 0.10$)e+43 \\
      N1550 & 0.0123 &  1.159 & $ 1.44^{+ 0.03}_{-0.02}$ & $   4.24\pm 0.37$  & ($ 6.93\pm 0.61$)e+42 & ($ 7.00\pm 0.62$)e+42 \\
      A2063 & 0.0354 &  0.292 & $ 3.56^{+ 0.16}_{-0.12}$ & $   4.23\pm 0.09$  & ($ 5.74\pm 0.12$)e+43 & ($ 5.90\pm 0.12$)e+43 \\
      A1644 & 0.0474 &  0.533 & $^{\natural} 4.70^{+ 0.90}_{-0.70}$ & $   4.09\pm 0.34$  & ($ 1.00\pm 0.08$)e+44 & ($ 1.04\pm 0.09$)e+44 \\
      A0119 & 0.0440 &  0.310 & $ 5.69^{+ 0.24}_{-0.28}$ & $   4.05\pm 0.06$  & ($ 8.53\pm 0.13$)e+43 & ($ 8.82\pm 0.13$)e+43 \\
 $^{*}$A0644 & 0.0704 &  0.514 & $ 6.54^{+ 0.27}_{-0.26}$ & $   3.97\pm 0.07$  & ($ 2.16\pm 0.04$)e+44 & ($ 2.28\pm 0.04$)e+44 \\
      N4636 & 0.0037 &  0.175 & $ 0.66^{+ 0.03}_{-0.01}$ & $   3.97\pm 0.47$  & ($ 5.84\pm 0.69$)e+41 & ($ 5.86\pm 0.69$)e+41 \\
      A3158 & 0.0590 &  0.106 & $ 5.41^{+ 0.26}_{-0.24}$ & $   3.79\pm 0.09$  & ($ 1.44\pm 0.04$)e+44 & ($ 1.51\pm 0.04$)e+44 \\
      A1736 & 0.0461 &  0.536 & $ 3.68^{+ 0.22}_{-0.17}$ & $   3.54\pm 0.36$  & ($ 8.21\pm 0.84$)e+43 & ($ 8.51\pm 0.87$)e+43 \\
      A0754 & 0.0528 &  0.459 & $ 9.00^{+ 0.35}_{-0.34}$ & $   3.34\pm 0.09$  & ($ 1.01\pm 0.03$)e+44 & ($ 1.06\pm 0.03$)e+44 \\
      A0399 & 0.0715 &  1.058 & $ 6.46^{+ 0.38}_{-0.36}$ & $   3.30\pm 0.29$  & ($ 1.85\pm 0.16$)e+44 & ($ 1.95\pm 0.17$)e+44 \\
      MKW3S & 0.0450 &  0.315 & $ 3.45^{+ 0.13}_{-0.10}$ & $   3.22\pm 0.05$  & ($ 7.13\pm 0.12$)e+43 & ($ 7.38\pm 0.12$)e+43 \\
 $^{*}$A0539 & 0.0288 &  1.206 & $ 3.04^{+ 0.11}_{-0.10}$ & $   3.12\pm 0.07$  & ($ 2.80\pm 0.06$)e+43 & ($ 2.87\pm 0.06$)e+43 \\
    EXO0422 & 0.0390 &  0.640 & $^{\natural} 2.90^{+ 0.90}_{-0.60}$ & $   3.12\pm 0.32$  & ($ 5.16\pm 0.52$)e+43 & ($ 5.32\pm 0.54$)e+43 \\
      A4059 & 0.0460 &  0.110 & $ 3.94^{+ 0.15}_{-0.15}$ & $   3.10\pm 0.07$  & ($ 7.16\pm 0.16$)e+43 & ($ 7.42\pm 0.16$)e+43 \\
      A3581 & 0.0214 &  0.426 & $ 1.83^{+ 0.04}_{-0.02}$ & $   3.08\pm 0.16$  & ($ 1.53\pm 0.08$)e+43 & ($ 1.55\pm 0.08$)e+43 \\
      A3112 & 0.0750 &  0.253 & $ 4.72^{+ 0.37}_{-0.25}$ & $   3.07\pm 0.06$  & ($ 1.90\pm 0.03$)e+44 & ($ 2.01\pm 0.04$)e+44 \\
      A0576 & 0.0381 &  0.569 & $ 3.83^{+ 0.16}_{-0.15}$ & $   2.99\pm 0.33$  & ($ 4.73\pm 0.53$)e+43 & ($ 4.88\pm 0.54$)e+43 \\
      A3562 & 0.0499 &  0.391 & $ 4.47^{+ 0.23}_{-0.21}$ & $   2.89\pm 0.04$  & ($ 7.87\pm 0.12$)e+43 & ($ 8.18\pm 0.13$)e+43 \\
      A2204 & 0.1523 &  0.594 & $ 6.38^{+ 0.23}_{-0.23}$ & $   2.73\pm 0.07$  & ($ 7.13\pm 0.19$)e+44 & ($ 7.92\pm 0.21$)e+44 \\
      A0400 & 0.0240 &  0.938 & $ 2.43^{+ 0.13}_{-0.12}$ & $   2.64\pm 0.05$  & ($ 1.65\pm 0.03$)e+43 & ($ 1.68\pm 0.03$)e+43 \\
\hline %------------------------------------------------
\end{tabular}
\end{center}
\end{table*}
\addtocounter{table}{-1}

\begin{table*}
\begin{center}
\caption{continued}
\begin{tabular}{cccrccc}
\hline %------------------------------------------------
\hline %------------------------------------------------
 NAME & redshift & $n_{\rm H}$
 & $kT_{\rm h}\ \ \ \ $
 & Flux(0.1-2.4keV)
 & \multicolumn{2}{c}{$^{\dag}$$L_{\rm X}$ (0.1-2.4keV) ($h^{-2}$ ergs s$^{-1}$ cm$^{-2}$)}\\
 & & ($10^{21}$cm$^{-2}$) & (keV)\ \ \ \ & ($10^{-11}$ergs s$^{-1}$ cm$^{-2}$) & Open & Flat \\ 
\hline %------------------------------------------------
      A2065 & 0.0721 &  0.284 & $ 5.37^{+ 0.34}_{-0.30}$ & $   2.55\pm 0.26$  & ($ 1.46\pm 0.15$)e+44 & ($ 1.54\pm 0.15$)e+44 \\
      A1651 & 0.0860 &  0.171 & $ 6.22^{+ 0.45}_{-0.41}$ & $   2.52\pm 0.05$  & ($ 2.06\pm 0.04$)e+44 & ($ 2.19\pm 0.04$)e+44 \\
      A2589 & 0.0416 &  0.439 & $ 3.38^{+ 0.13}_{-0.13}$ & $   2.52\pm 0.05$  & ($ 4.75\pm 0.10$)e+43 & ($ 4.91\pm 0.11$)e+43 \\
       MKW8 & 0.0270 &  0.260 & $ 3.29^{+ 0.23}_{-0.22}$ & $   2.51\pm 0.35$  & ($ 1.99\pm 0.27$)e+43 & ($ 2.03\pm 0.28$)e+43 \\
      A2657 & 0.0404 &  0.527 & $ 3.53^{+ 0.12}_{-0.12}$ & $   2.49\pm 0.04$  & ($ 4.44\pm 0.06$)e+43 & ($ 4.58\pm 0.07$)e+43 \\
      A3376 & 0.0455 &  0.501 & $ 4.43^{+ 0.39}_{-0.38}$ & $   2.44\pm 0.06$  & ($ 5.52\pm 0.13$)e+43 & ($ 5.72\pm 0.14$)e+43 \\
      S1101 & 0.0580 &  0.185 & $^{\natural} 2.60^{+ 0.50}_{-0.50}$ & $   2.44\pm 0.04$  & ($ 9.05\pm 0.14$)e+43 & ($ 9.46\pm 0.14$)e+43 \\
      A1650 & 0.0845 &  0.154 & $ 5.68^{+ 0.30}_{-0.27}$ & $   2.43\pm 0.26$  & ($ 1.92\pm 0.21$)e+44 & ($ 2.04\pm 0.22$)e+44 \\
      A2634 & 0.0312 &  0.517 & $ 3.45^{+ 0.16}_{-0.16}$ & $   2.38\pm 0.06$  & ($ 2.51\pm 0.06$)e+43 & ($ 2.58\pm 0.07$)e+43 \\
      A3391 & 0.0531 &  0.542 & $ 5.89^{+ 0.45}_{-0.33}$ & $   2.20\pm 0.07$  & ($ 6.78\pm 0.21$)e+43 & ($ 7.06\pm 0.22$)e+43 \\
      A2597 & 0.0852 &  0.250 & $ 4.20^{+ 0.49}_{-0.41}$ & $   2.20\pm 0.04$  & ($ 1.77\pm 0.04$)e+44 & ($ 1.88\pm 0.04$)e+44 \\
   ZwCl1215 & 0.0750 &  0.164 & $^{\flat} 6.30^{+ 2.68}_{-1.88}$ & $   2.17\pm 0.05$  & ($ 1.34\pm 0.03$)e+44 & ($ 1.42\pm 0.03$)e+44 \\
           &        &       & $^{\sharp} 6.36^{+ 2.94}_{-2.01}$ &  &  &  \\
      A2244 & 0.0970 &  0.207 & $ 5.77^{+ 0.61}_{-0.44}$ & $   2.10\pm 0.07$  & ($ 2.19\pm 0.08$)e+44 & ($ 2.35\pm 0.08$)e+44 \\
      A0133 & 0.0569 &  0.160 & $ 3.97^{+ 0.28}_{-0.27}$ & $   2.06\pm 0.03$  & ($ 7.33\pm 0.10$)e+43 & ($ 7.66\pm 0.11$)e+43 \\
      A2163 & 0.2010 &  1.227 & $10.55^{+ 1.01}_{-0.68}$ & $   2.04\pm 0.05$  & ($ 9.28\pm 0.23$)e+44 & ($10.59\pm 0.27$)e+44 \\
      A2255 & 0.0800 &  0.251 & $ 5.92^{+ 0.40}_{-0.26}$ & $   2.02\pm 0.04$  & ($ 1.42\pm 0.03$)e+44 & ($ 1.51\pm 0.03$)e+44 \\
    IIIZw54 & 0.0311 &  1.668 & $^{\flat} 2.98^{+ 1.27}_{-0.89}$ & $   2.01\pm 0.25$  & ($ 2.11\pm 0.27$)e+43 & ($ 2.16\pm 0.27$)e+43 \\
           &        &       & $^{\sharp} 3.00^{+ 1.39}_{-0.95}$ &  &  &  \\
     A3395s & 0.0498 &  0.849 & $ 5.55^{+ 0.89}_{-0.65}$ & $   2.01\pm 0.13$  & ($ 5.42\pm 0.34$)e+43 & ($ 5.63\pm 0.35$)e+43 \\
       N507 & 0.0165 &  0.525 & $ 1.40^{+ 0.04}_{-0.07}$ & $   2.00\pm 0.04$  & ($ 5.88\pm 0.12$)e+42 & ($ 5.96\pm 0.12$)e+42 \\
       MKW4 & 0.0200 &  0.186 & $ 1.84^{+ 0.05}_{-0.03}$ & $   2.00\pm 0.05$  & ($ 8.65\pm 0.24$)e+42 & ($ 8.79\pm 0.24$)e+42 \\
   UGC03957 & 0.0340 &  0.459 & $^{\flat} 3.19^{+ 1.36}_{-0.95}$ & $   1.98\pm 0.19$  & ($ 2.49\pm 0.24$)e+43 & ($ 2.56\pm 0.25$)e+43 \\
           &        &       & $^{\sharp} 3.21^{+ 1.48}_{-1.02}$ &  &  &  \\
      A3827 & 0.0980 &  0.284 & $^{\flat} 7.55^{+ 3.21}_{-2.25}$ & $   1.98\pm 0.19$  & ($ 2.10\pm 0.20$)e+44 & ($ 2.25\pm 0.22$)e+44 \\
           &        &       & $^{\sharp} 7.66^{+ 3.54}_{-2.42}$ &  &  &  \\
      A3822 & 0.0760 &  0.212 & $ 5.12^{+ 0.43}_{-0.31}$ & $   1.98\pm 0.24$  & ($ 1.26\pm 0.15$)e+44 & ($ 1.33\pm 0.16$)e+44 \\
    IIZw108 & 0.0494 &  0.663 & $^{\flat} 4.25^{+ 1.81}_{-1.27}$ & $   1.90\pm 0.23$  & ($ 5.06\pm 0.61$)e+43 & ($ 5.26\pm 0.63$)e+43 \\
           &        &       & $^{\sharp} 4.28^{+ 1.98}_{-1.35}$ &  &  &  \\
        M49 & 0.0044 &  0.159 & $ 1.33^{+ 0.03}_{-0.03}$ & $   1.89\pm 0.03$  & ($ 3.93\pm 0.07$)e+41 & ($ 3.94\pm 0.07$)e+41 \\
   ZwCl1742 & 0.0757 &  0.356 & $^{\flat} 5.99^{+ 2.55}_{-1.79}$ & $   1.87\pm 0.13$  & ($ 1.18\pm 0.08$)e+44 & ($ 1.25\pm 0.09$)e+44 \\
           &        &       & $^{\sharp} 6.05^{+ 2.80}_{-1.91}$ &  &  &  \\
       S405 & 0.0613 &  0.765 & $^{\flat} 4.98^{+ 2.12}_{-1.49}$ & $   1.82\pm 0.24$  & ($ 7.48\pm 1.00$)e+43 & ($ 7.84\pm 1.05$)e+43 \\
           &        &       & $^{\sharp} 5.02^{+ 2.32}_{-1.59}$ &  &  &  \\
      A3532 & 0.0539 &  0.596 & $ 4.41^{+ 0.19}_{-0.18}$ & $   1.78\pm 0.05$  & ($ 5.65\pm 0.17$)e+43 & ($ 5.88\pm 0.17$)e+43 \\
      A3695 & 0.0890 &  0.356 & $^{\flat} 6.67^{+ 2.84}_{-1.99}$ & $   1.76\pm 0.26$  & ($ 1.54\pm 0.23$)e+44 & ($ 1.65\pm 0.25$)e+44 \\
           &        &       & $^{\sharp} 6.76^{+ 3.12}_{-2.14}$ &  &  &  \\
      HCG94 & 0.0417 &  0.455 & $ 3.30^{+ 0.17}_{-0.16}$ & $   1.72\pm 0.03$  & ($ 3.25\pm 0.05$)e+43 & ($ 3.36\pm 0.06$)e+43 \\
     A3528s & 0.0551 &  0.610 & $ 4.60^{+ 0.49}_{-0.27}$ & $   1.70\pm 0.04$  & ($ 5.65\pm 0.15$)e+43 & ($ 5.89\pm 0.15$)e+43 \\
       S540 & 0.0358 &  0.353 & $^{\flat} 3.07^{+ 1.30}_{-0.91}$ & $   1.62\pm 0.13$  & ($ 2.26\pm 0.18$)e+43 & ($ 2.32\pm 0.19$)e+43 \\
           &        &       & $^{\sharp} 3.09^{+ 1.43}_{-0.98}$ &  &  &  \\
      A2877 & 0.0241 &  0.210 & $^{\natural} 3.50^{+ 2.20}_{-1.10}$ & $   1.61\pm 0.03$  & ($ 1.01\pm 0.02$)e+43 & ($ 1.03\pm 0.02$)e+43 \\
     A3395n & 0.0498 &  0.542 & $ 5.11^{+ 0.47}_{-0.43}$ & $   1.54\pm 0.10$  & ($ 4.16\pm 0.27$)e+43 & ($ 4.33\pm 0.28$)e+43 \\
     A2151w & 0.0369 &  0.336 & $ 2.58^{+ 0.19}_{-0.20}$ & $   1.53\pm 0.05$  & ($ 2.26\pm 0.07$)e+43 & ($ 2.33\pm 0.07$)e+43 \\
      A3560 & 0.0495 &  0.392 & $^{\flat} 3.87^{+ 1.65}_{-1.15}$ & $   1.50\pm 0.06$  & ($ 4.02\pm 0.17$)e+43 & ($ 4.17\pm 0.17$)e+43 \\
           &        &       & $^{\sharp} 3.90^{+ 1.81}_{-1.23}$ &  &  &  \\
      A2734 & 0.0620 &  0.184 & $ 5.07^{+ 0.36}_{-0.42}$ & $   1.47\pm 0.06$  & ($ 6.19\pm 0.25$)e+43 & ($ 6.49\pm 0.26$)e+43 \\
     A0548e & 0.0410 &  0.188 & $ 2.93^{+ 0.17}_{-0.15}$ & $   1.46\pm 0.04$  & ($ 2.68\pm 0.08$)e+43 & ($ 2.77\pm 0.08$)e+43 \\
      A1689 & 0.1840 &  0.180 & $ 8.58^{+ 0.84}_{-0.40}$ & $   1.45\pm 0.03$  & ($ 5.53\pm 0.10$)e+44 & ($ 6.25\pm 0.11$)e+44 \\
      A1914 & 0.1712 &  0.097 & $ 8.41^{+ 0.60}_{-0.58}$ & $   1.45\pm 0.03$  & ($ 4.77\pm 0.11$)e+44 & ($ 5.36\pm 0.12$)e+44 \\
    RXJ2344 & 0.0786 &  0.354 & $^{\flat} 5.45^{+ 2.32}_{-1.62}$ & $   1.37\pm 0.03$  & ($ 9.34\pm 0.21$)e+43 & ($ 9.91\pm 0.23$)e+43 \\
           &        &       & $^{\sharp} 5.52^{+ 2.55}_{-1.74}$ &  &  &  \\
      A3921 & 0.0936 &  0.280 & $ 5.39^{+ 0.38}_{-0.35}$ & $   1.31\pm 0.04$  & ($ 1.27\pm 0.03$)e+44 & ($ 1.36\pm 0.04$)e+44 \\
      A1413 & 0.1427 &  0.162 & $ 6.56^{+ 0.65}_{-0.44}$ & $   1.28\pm 0.03$  & ($ 2.92\pm 0.08$)e+44 & ($ 3.23\pm 0.09$)e+44 \\
\hline %------------------------------------------------
\end{tabular}
\end{center}
\end{table*}
\addtocounter{table}{-1}

\begin{table*}
\begin{center}
\caption{continued}
\begin{tabular}{cccrccc}
\hline %------------------------------------------------
\hline %------------------------------------------------
 NAME & redshift & $n_{\rm H}$
 & $kT_{\rm h}\ \ \ \ $
 & Flux(0.1-2.4keV)
 & \multicolumn{2}{c}{$^{\dag}$$L_{\rm X}$ (0.1-2.4keV) ($h^{-2}$ ergs s$^{-1}$ cm$^{-2}$)}\\
 & & ($10^{21}$cm$^{-2}$) & (keV)\ \ \ \ & ($10^{-11}$ergs s$^{-1}$ cm$^{-2}$) & Open & Flat \\ 
\hline %------------------------------------------------
      N5813 & 0.0064 &  0.419 & $ 0.76^{+ 0.19}_{-0.19}$ & $   1.27\pm 0.14$  & ($ 5.62\pm 0.62$)e+41 & ($ 5.66\pm 0.62$)e+41 \\
      A1775 & 0.0757 &  0.100 & $ 3.66^{+ 0.34}_{-0.20}$ & $   1.26\pm 0.04$  & ($ 8.02\pm 0.23$)e+43 & ($ 8.49\pm 0.25$)e+43 \\
     A3528n & 0.0540 &  0.610 & $ 4.79^{+ 0.50}_{-0.44}$ & $   1.26\pm 0.05$  & ($ 4.00\pm 0.15$)e+43 & ($ 4.17\pm 0.16$)e+43 \\
      A1800 & 0.0748 &  0.118 & $^{\flat} 4.95^{+ 2.11}_{-1.48}$ & $   1.20\pm 0.15$  & ($ 7.39\pm 0.95$)e+43 & ($ 7.82\pm 1.01$)e+43 \\
           &        &       & $^{\sharp} 5.02^{+ 2.32}_{-1.59}$ &  &  &  \\
      A3888 & 0.1510 &  0.120 & $^{\flat} 8.46^{+ 3.60}_{-2.53}$ & $   1.09\pm 0.04$  & ($ 2.77\pm 0.11$)e+44 & ($ 3.08\pm 0.12$)e+44 \\
           &        &       & $^{\sharp} 8.68^{+ 4.01}_{-2.75}$ &  &  &  \\
      A3530 & 0.0544 &  0.600 & $ 4.05^{+ 0.32}_{-0.30}$ & $   0.96\pm 0.04$  & ($ 3.11\pm 0.15$)e+43 & ($ 3.25\pm 0.15$)e+43 \\
      N5846 & 0.0061 &  0.425 & $ 0.64^{+ 0.04}_{-0.03}$ & $   0.83\pm 0.03$  & ($ 3.35\pm 0.12$)e+41 & ($ 3.36\pm 0.12$)e+41 \\
       N499 & 0.0147 &  0.525 & $ 0.66^{+ 0.02}_{-0.03}$ & $   0.45\pm 0.02$  & ($ 1.07\pm 0.04$)e+42 & ($ 1.08\pm 0.05$)e+42 \\
     A0548w & 0.0424 &  0.179 & $^{\flat} 1.64^{+ 0.70}_{-0.49}$ & $   0.25\pm 0.02$  & ($ 4.84\pm 0.43$)e+42 & ($ 5.03\pm 0.45$)e+42 \\
           &        &       & $^{\sharp} 1.68^{+ 0.77}_{-0.53}$ &  &  &  \\
\hline %------------------------------------------------
\end{tabular}
\end{center}
\begin{itemize}
\setlength{\itemsep}{2mm}
\item[$^{*)}$]: The Galactic latitude is lower than 20$^{\circ}$.
\item[$^{\dag)}$]: The values are calculated for
an open ($\Omega_{\rm m,0}=0.2$, $\Omega_{\Lambda,0}=0$) 
and a flat ($\Omega_{\rm m,0}=0.2$, $\Omega_{\Lambda,0}=0.8$) universe.
\item[$^{\flat)}$]: The temperature is estimated with the $L-T$ relation assuming
the open universe.
\item[$^{\sharp)}$]: The temperature is estimated with the $L-T$ relation assuming
the flat universe.
\item[$^{\natural)}$]: The temperature is derived with non-{\it ASCA} spectroscopy.
{\it Einstein} MPC for A2877 and A1644 (David et al. \cite{David93}),
{\it EXOSAT} LE+ME for EXO0422 (Edge \& Stewart \cite{EdgeStewart91}),
and {\it XMM-Newton} EPIC for S1101 (Kaastra et al. \cite{Kaastra01}).
\end{itemize}
\label{allsample}
\end{table*}

%----2.3---2.3---2.3---2.3---2.3---2.3---2.3---2.3---2.3----
\subsection{$L-T$ relation}
%----2.3---2.3---2.3---2.3---2.3---2.3---2.3---2.3---2.3----
%-- Calculation --\\
For the 88 clusters observed with {\it ASCA},
X-ray fluxes and luminosities were calculated
from the spectral-analysis results described above
and the {\it ROSAT} PSPC count rates in 50-201 PI channel
given by Reiprich \& B\"{o}hringer (\cite{RB}).
Assuming isothermality in each cluster at the hot-component temperature
and the Galactic column density given in Table 1,
we determined the 0.1--2.4~keV fluxes and luminosities that reproduce
the PSPC count rates.
As summarized in Table 1, the luminosities were calculated
for two different cosmological models 
with an open ($\Omega_{\rm m,0}=0.2, \Omega_{\Lambda,0}=0$)
and a flat ($\Omega_{\rm m,0}=0.2, \Omega_{\Lambda,0}=0.8$) universe.

\begin{figure}
\centerline{
\psfig{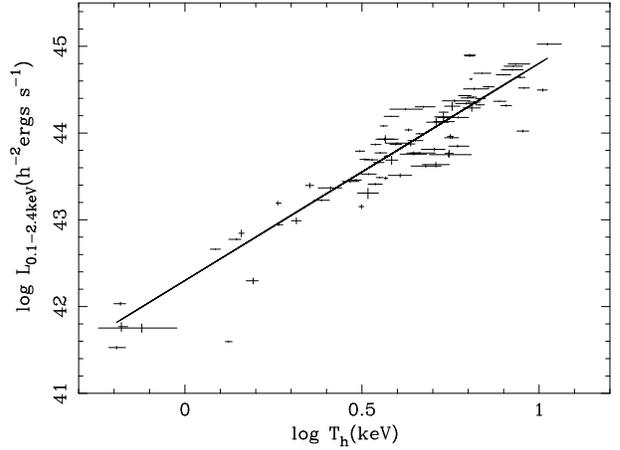}}
\caption{The luminosities in the energy range of 0.1-2.4~keV in the cluster
rest frame in a flat universe 
($\Omega_{\rm m,0}=0.2$, $\Omega_{\Lambda,0}=0.8$)
as a function of the hot component temperature 
obtained from the two-temperature model fitted to the {\it ASCA} data. 
The best-fit power-law function (eq. \ref{LTflat}) is overlayed.}
\label{L-T}
\end{figure}

%-- L-T relation --\\
We then established a correlation between the luminosity
in the 0.1--2.4~keV band, $L_{\rm 0.1-2.4keV}$,
and the temperature of the hot component, $T_{\rm h}$,
as illustrated in Fig.~\ref{L-T}.
The $L_{\rm 0.1-2.4keV}$-$T_{\rm h}$ relations for
the open and flat universe were fitted individually
with power-law functions by a linear regression method.
For the fit we did not use clusters with a hot component temperature
cooler than 1.4~keV,
because it is known that
a single power-law function is not a good representation
of the $L-T$ relation over a wide temperature range.
In particular, including clusters, groups, and elliptical galaxies,
the $L-T$ relation becomes steeper below $\sim1$~keV
(e.g. Xue \& Wu \cite{XW}).
Moreover, the theoretically predicted XTFs as given in Sect. 4 may not be
directly compared with an observed XTF (see Sect. 4.3 for more explanation).
Therefore, in the power-law fitting, we exclude 6 sample clusters
cooler than 1.4~keV and use only 82 clusters.
For the linear regression fit,
we employed the BCES(X$_{2}$$|$X$_{1}$) estimator 
given by Akritas \& Bershady (\cite{AB})
(In this case, the luminosity and
temperature are assigned as X$_2$ and X$_1$, respectively).
The choice of the estimator is based on an argument by
Isobe et al. (\cite{Isobe}).
The best-fit functions are
\begin{eqnarray}
\log L_{\rm 0.1-2.4keV}(h^{-2}{\rm ergs\ s}^{-1})=\hspace{3cm} & & \nonumber\\
 42.31\ (\pm 0.09) + 2.47\ (\pm0.14)\ \log T_{\rm h}({\rm keV}) & &
\label{LTopen}
\end{eqnarray}
for the open universe,
and 
\begin{eqnarray}
\log L_{\rm 0.1-2.4keV}(h^{-2}{\rm ergs\ s}^{-1})=\hspace{3cm} & & \nonumber\\
 42.30\ (\pm 0.09) + 2.50\ (\pm0.14)\ \log T_{\rm h}({\rm keV}) & &
\label{LTflat}
\end{eqnarray}
for the flat universe.

%-- a correction on the normalization
We have made a correction
to the derived $L_{\rm 0.1-2.4keV}-T_{\rm h}$ relation.
Since the less luminous clusters are less likely to be sampled,
the apparent {\it mean} luminosity at a given temperature,
which is estimated by the simple regression fit,
is biased towards higher values than the true mean value.
As will be shown in Sect. 4,
we studied this effect and corrected the normalization
to derive the ``true'' $L_{\rm 0.1-2.4keV}-T_{\rm h}$ relation,
as 
\begin{equation}
\log L_{\rm 0.1-2.4keV} (h^{-2} {\rm ergs\ s}^{-1})
= 42.15 + 2.47\ \log T_{\rm h}({\rm keV})
\label{revLTopen}
\end{equation}
for the open universe,
and 
\begin{equation}
\log L_{\rm 0.1-2.4keV} (h^{-2} {\rm ergs\ s}^{-1})
= 42.14 + 2.50\ \log T_{\rm h} ({\rm keV})
\label{revLTflat}
\end{equation}
for the flat universe.
The correction reduces the luminosity at a given temperature by 30\%.
The corrected $L_{\rm 0.1-2.4keV}-T_{\rm h}$ relation
will also be used to calculate $V_{\rm max}$ in Sect. 3.
The 30\% reduction of the mean temperature
results in a reduction of $V_{\rm max}$ (see Sect. 3) by 40\%.

%-- Temperature estimate from the L-T --\\
Using the corrected $L_{\rm 0.1-2.4keV}-T_{\rm h}$ relations,
we estimated the temperatures of 14 clusters for
which no {\it ASCA} data nor other spectroscopic measurement are available.
An $L_{\rm 0.1-2.4keV}$ and $T_{\rm h}$ combination was chosen
from the $L_{\rm 0.1-2.4keV}-T_{\rm h}$ relation
such that the predicted PSPC count rate equals the observed value.
Scatter around the best-fit power-law in the $L-T$ relation
was assigned as error.
The estimated temperatures and luminosities for the open and flat universe
cases as well as the corresponding 0.1--2.4~keV fluxes which are the same 
for the open and flat universe cases are summarized in Table 1.

%----2.4---2.4---2.4---2.4---2.4---2.4---2.4---2.4---2.4----
\subsection{Construction of the flux limited sample}
\begin{figure}
\centerline{
\psfig{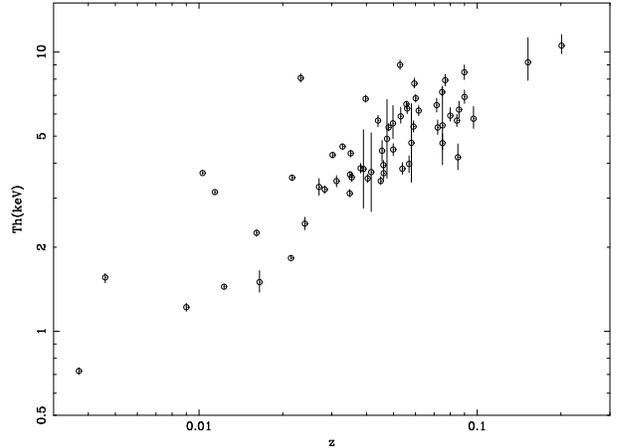}}
\caption{X-ray temperature vs redshift for the clusters of the flux-limited
sample.}
\label{T-z}
\end{figure}

%----2.4---2.4---2.4---2.4---2.4---2.4---2.4---2.4---2.4----
We set a flux-limit of $1.99\times10^{-11}$ ergs s$^{-1}$ cm$^{-2}$
in 0.1--2.4~keV band to construct a flux-limited complete sample
from the master catalog of 106 clusters.
As in Reiprich \& B\"{o}hringer (\cite{RB}) for {\sfsl HIFLUGCS},
additional selection criteria applied are that
the absolute Galactic latitude is greater than $20^{\circ}$,
and the cluster is located neither in the Magellanic Clouds
nor in the Virgo cluster region.
The total number of clusters thus selected is 63,
whose temperatures and redshifts are shown in Fig.~\ref{T-z}.
Since the temperatures of the individual clusters were newly determined
in this paper,
the fluxes estimated from the PSPC count rate
are slightly revised from the value in Reiprich \& B\"{o}hringer (\cite{RB})
where the temperatures are compiled from literature.
All of the 63 clusters selected here, however, still correspond with
the 63 {\sfsl HIFLUGCS} clusters.
Tests performed in Reiprich \& B\"{o}hringer (\cite{RB})
showed no indication of incompleteness of the sample.

In the following analysis, we exclude 2 clusters cooler than 1.4~keV
(see Sect. 2.3 and 4.3) from the flux-limited complete sample,
and use 61 clusters to study the XTF and make constraints on
cosmological parameters.
This is the largest complete sample of clusters up to now
with temperature measurements or reliable temperature estimates.
We also stress the accuracy of the temperature measurements for
the sample clusters:
Among the 61 clusters used for constructing the XTF,
the temperatures of 56 clusters (90\% of the sample) 
were measured with {\it ASCA} data with 90\% errors of 6-10\%,
other spectroscopic measurements were used for 3 clusters,
and there are only two clusters for which the temperature is
estimated with the $L-T$ relation.

%=====3=====3=====3=====3=====3=====3=====3=====3=====3=====
\section{The Temperature Function}
%=====3=====3=====3=====3=====3=====3=====3=====3=====3=====
\subsection{Derivation of XTF}
\begin{figure}
\centerline{
\psfig{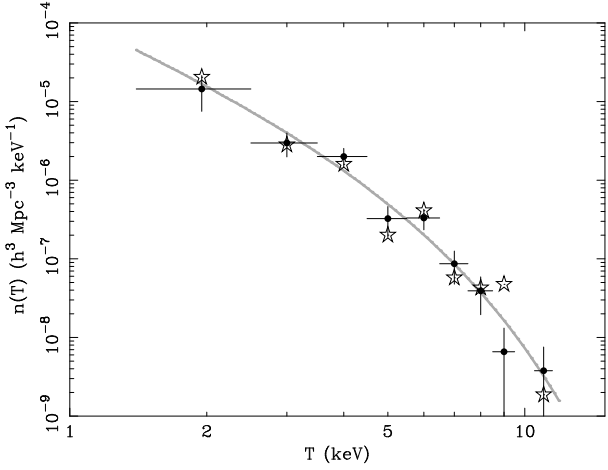}}
\caption{The X-ray temperature function derived with
the flux-limited complete sample of 61 clusters
for the case of a flat universe 
($\Omega_{\rm m,0}=0.2, \Omega_{\Lambda,0}=0.8$).
Filled circles show the XTF evaluated with $1/V_{\rm max}(T_{\rm i})$, while
open stars show the one obtained with $1/v_{\rm max}(L_{\rm i},T_{\rm i})$.
The temperature bin widths are 1~keV, except for the lowest temperature
bin that includes $kT$=1.4--2.5~keV.
The vertical error bars for filled circles indicate Poisson errors.
Open stars should have the same Poisson errors, which are omitted
for a clear display.
The solid curve is the best-fit Press-Schechter function derived in Sect. 4.1.}
\label{XTF+SIMPLE}
\end{figure}

Using the X-ray flux-limited complete sample of 61 clusters
($T_{\rm h} \geq 1.4$),
we address in this section the XTF describing
the number density of clusters of galaxies in comoving
coordinates at present as a function of the X-ray temperature.
A full description of the analysis method can be found in the Appendix.
However, the key points are presented here.
The differential temperature function, $\phi(T)$, is defined as
the number density of clusters in the temperature range
from $T-\frac{d T}{2}$ to $T+\frac{d T}{2}$.
From a finite number of sample clusters,
it can be evaluated as
\begin{equation}
\phi(T)dT = \sum_{T-\frac{\Delta T}{2}\le T_{\rm i}<T+\frac{\Delta T}{2}}
	\frac{1}{V_{\rm max}(T_{\rm i})}
\label{eq:XTF}
\end{equation}
where $i$ denotes a sequential number of clusters in a temperature
range from $T-\frac{\Delta T}{2}$ to $T+\frac{\Delta T}{2}$,
and $V_{\rm max}(T_{\rm i})$ is a maximum comoving volume in which
a cluster having a temperature of $T_{\rm i}$ could have been detected
under the selection criteria given above.
Following Markevitch (\cite{Markevitch}),
we assume, for a given temperature, the logarithmic luminosity, log $L$,
follows a Gaussian distribution around the mean
which is given with the $L-T$ relation, i.e. $L=AT^{\alpha}$,
and with a constant standard deviation of $\sigma_{\log L}$.
$V_{\rm max}(T)$ then can be evaluated as
\begin{eqnarray}
V_{\rm max}(T) = \hspace{7cm} & & \nonumber\\
 \int^{\infty}_{-\infty} 
 \frac{v_{\rm max}(L,T)}{\sqrt{2\pi \sigma_{\log L}^2}}
 \exp\left\{ -\frac{(\log AT^{\alpha}-\log L)^2}{2\sigma_{\log L}^2} \right\}
 d\log L,\ \ & &
\label{eq:VmaxT}
\end{eqnarray}
where $v_{\rm max}(L,T)$ is the maximum search volume for a cluster
having the luminosity of $L$ and the temperature of $T$
and is given by Eq. \ref{eq:modvmax}.
Recall that $V_{\rm max}(T)$ depends on the cosmological parameters
and was calculated for an open ($\Omega_{\rm m,0}=0.2, \Omega_{\Lambda,0}=0$)
and a flat ($\Omega_{\rm m,0}=0.2, \Omega_{\Lambda,0}=0.8$) universe.
The results are listed for individual clusters in Table 2.
The XTF thus derived with 1~keV bin widths is illustrated
in Fig.~\ref{XTF+SIMPLE}.

The improvement of the new result over previous work
is mainly characterized by a significant broadening of the temperature range
towards lower temperature systems down to 1.4~keV.
A single power law is no longer a good description to the observationally
measured XTF and an exponential cut off towards the higher temperature
is clearly seen, as a theoretical model predicts.
The detection of such a curvature brings tight constraints
on cosmological parameters as will be shown in the next section.

Conventionally, $V_{\rm max}(T_{\rm i})$ in Eq. \ref{eq:XTF}
is often simply replaced by $v_{\rm max}(L_{\rm i},T_{\rm i})$
to evaluate the XTF (e.g. Henry \cite{Henry}).
For comparison, the XTF evaluated with 
$\phi(T)dT = \sum 1/v_{\rm max}(L_{\rm i},T_{\rm i})$
is also overlayed in Fig.~\ref{XTF+SIMPLE}, 
which overall shows good agreement
with the one with $\sum 1/V_{\rm max}(T_{\rm i})$,
but exhibits larger variance
due to the intrinsic scatter in the $L-T$ relation.

\begin{table}
\begin{center}
\caption{Summary of the flux-limited complete sample}
\begin{tabular}{clcc}
\hline %------------------------------------------------
\hline %------------------------------------------------
 NAME & $kT_{\rm h}$(keV) & 
 \multicolumn{2}{c}{$^{\dag)}V_{\rm max}$($h^{-3}$ Mpc$^{-3}$)} \\
      &                      & Open & Flat \\
\hline %------------------------------------------------
     A2163 &$^{\ }$ 10.55 &   2.39e+08 &   2.65e+08\\
     A0754 &$^{\ }$  9.00 &   1.40e+08 &   1.52e+08\\
     A2142 &$^{\ }$  8.46 &   1.14e+08 &   1.23e+08\\
      COMA &$^{\ }$  8.07 &   9.69e+07 &   1.04e+08\\
     A2029 &$^{\ }$  7.93 &   9.12e+07 &   9.78e+07\\
     A3266 &$^{\ }$  7.72 &   8.30e+07 &   8.89e+07\\
     A0401 &$^{\ }$  7.19 &   6.50e+07 &   6.91e+07\\
     A0478 &$^{\ }$  6.91 &   5.66e+07 &   6.00e+07\\
     A2256 &$^{\ }$  6.83 &   5.45e+07 &   5.77e+07\\
     A3571 &$^{\ }$  6.80 &   5.36e+07 &   5.67e+07\\
     A0085 &$^{\ }$  6.51 &   4.60e+07 &   4.87e+07\\
     A0399 &$^{\ }$  6.46 &   4.46e+07 &   4.72e+07\\
     A2204 &$^{\ }$  6.38 &   4.28e+07 &   4.52e+07\\
  ZwCl1215 & $^{\flat}$  6.30 \ \ $^{\sharp}$  6.36 &   4.09e+07 &   4.49e+07\\
     A3667 &$^{\ }$  6.28 &   4.05e+07 &   4.28e+07\\
     A1651 &$^{\ }$  6.22 &   3.92e+07 &   4.14e+07\\
     A1795 &$^{\ }$  6.17 &   3.81e+07 &   4.02e+07\\
     A2255 &$^{\ }$  5.92 &   3.29e+07 &   3.46e+07\\
     A3391 &$^{\ }$  5.89 &   3.23e+07 &   3.40e+07\\
     A2244 &$^{\ }$  5.77 &   3.01e+07 &   3.17e+07\\
     A0119 &$^{\ }$  5.69 &   2.86e+07 &   3.00e+07\\
     A1650 &$^{\ }$  5.68 &   2.85e+07 &   2.99e+07\\
    A3395s &$^{\ }$  5.55 &   2.64e+07 &   2.76e+07\\
     A3158 &$^{\ }$  5.41 &   2.40e+07 &   2.50e+07\\
     A2065 &$^{\ }$  5.37 &   2.34e+07 &   2.44e+07\\
     A3558 &$^{\ }$  5.37 &   2.35e+07 &   2.45e+07\\
     A3112 &$^{\ }$  4.72 &   1.48e+07 &   1.53e+07\\
     A1644 & $^{\natural}$  4.70 &   1.46e+07 &   1.51e+07\\
     A0496 &$^{\ }$  4.59 &   1.34e+07 &   1.39e+07\\
     A3562 &$^{\ }$  4.47 &   1.22e+07 &   1.27e+07\\
     A3376 &$^{\ }$  4.43 &   1.18e+07 &   1.22e+07\\
     A2147 &$^{\ }$  4.34 &   1.10e+07 &   1.13e+07\\
     A2199 &$^{\ }$  4.28 &   1.04e+07 &   1.08e+07\\
     A2597 &$^{\ }$  4.20 &   9.75e+06 &   1.01e+07\\
     A0133 &$^{\ }$  3.97 &   8.02e+06 &   8.25e+06\\
     A4059 &$^{\ }$  3.94 &   7.76e+06 &   7.98e+06\\
     A0576 &$^{\ }$  3.83 &   7.07e+06 &   7.26e+06\\
   HYDRA-A &$^{\ }$  3.82 &   6.98e+06 &   7.16e+06\\
     A3526 &$^{\ }$  3.69 &   6.15e+06 &   6.30e+06\\
     A1736 &$^{\ }$  3.68 &   6.11e+06 &   6.26e+06\\
    2A0335 &$^{\ }$  3.64 &   5.87e+06 &   6.00e+06\\
     A2063 &$^{\ }$  3.56 &   5.41e+06 &   5.53e+06\\
     A1367 &$^{\ }$  3.55 &   5.36e+06 &   5.48e+06\\
     A2657 &$^{\ }$  3.53 &   5.25e+06 &   5.36e+06\\
     A2634 &$^{\ }$  3.45 &   4.84e+06 &   4.94e+06\\
     MKW3S &$^{\ }$  3.45 &   4.87e+06 &   4.97e+06\\
     A2589 &$^{\ }$  3.38 &   4.48e+06 &   4.57e+06\\
      MKW8 &$^{\ }$  3.29 &   4.08e+06 &   4.16e+06\\
     A4038 &$^{\ }$  3.22 &   3.78e+06 &   3.85e+06\\
     A1060 &$^{\ }$  3.15 &   3.50e+06 &   3.56e+06\\
     A2052 &$^{\ }$  3.12 &   3.37e+06 &   3.43e+06\\
   IIIZw54 & $^{\flat}$  2.98 \ \ $^{\sharp}$  3.00 &   2.87e+06 &   2.99e+06\\
   EXO0422 & $^{\natural}$  2.90 &   2.60e+06 &   2.63e+06\\
     S1101 & $^{\natural}$  2.60 &   1.76e+06 &   1.77e+06\\
     A0400 &$^{\ }$  2.43 &   1.37e+06 &   1.38e+06\\
\hline %------------------------------------------------
\end{tabular}
\end{center}
\end{table}
\addtocounter{table}{-1}
\begin{table}
\begin{center}
\caption{continued}
\begin{tabular}{clcc}
\hline %------------------------------------------------
\hline %------------------------------------------------
 NAME & $kT_{\rm h}$(keV) & 
 \multicolumn{2}{c}{$^{\dag)}V_{\rm max}$($h^{-3}$ Mpc$^{-3}$)} \\
      &                      & Open & Flat \\
\hline %------------------------------------------------
     A0262 &$^{\ }$  2.25 &   1.04e+06 &   1.04e+06\\
      MKW4 &$^{\ }$  1.84 &   5.00e+05 &   4.96e+05\\
     A3581 &$^{\ }$  1.83 &   4.88e+05 &   4.84e+05\\
    FORNAX &$^{\ }$  1.56 &   2.72e+05 &   2.68e+05\\
     N1550 &$^{\ }$  1.44 &   2.05e+05 &   2.02e+05\\
      N507 &$^{\ }$  1.40 &   1.83e+05 &   1.79e+05\\
     N5044 &$^{\ }$  1.22 &   1.10e+05 &   1.08e+05\\
     N4636 &$^{\ }$  0.66 &   1.16e+04 &   1.10e+04\\
\hline %------------------------------------------------
\end{tabular}
\end{center}
\begin{itemize}
\setlength{\itemsep}{2mm}
\item[$^{\dag)}$,$^{\flat)}$,$^{\sharp)}$,$^{\natural)}$]: as Table 1.
\end{itemize}
\label{FLS}
\end{table}

\subsection{Comparison to previous results}
\begin{figure}
\centerline{
\psfig{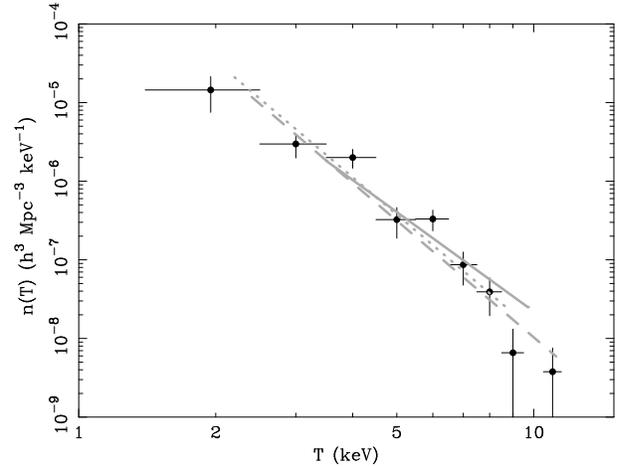}}
\caption{The X-ray temperature function derived with 
the flux-limited complete sample of 61 clusters
for the case of a flat universe
($\Omega_{\rm m,0}=0.2, \Omega_{\Lambda,0}=0.8$).
The three lines indicate previous measurements in the form of the best-fit
power-law functions. Solid, dotted, and dashed lines
correspond to the results by 
Markevitch (\cite{Markevitch}), 
Edge et al. (\cite{Edge}), 
and 
Henry (\cite{Henry}), 
respectively.
}
\label{XTF}
\end{figure}

In Fig.~\ref{XTF}, 
we compare our result to previously obtained XTFs overlayed
in the form of best-fit power law functions.
In the $3-8$~keV range, our XTF shows good agreement with
the previous results.

\begin{figure}
\psfig{file=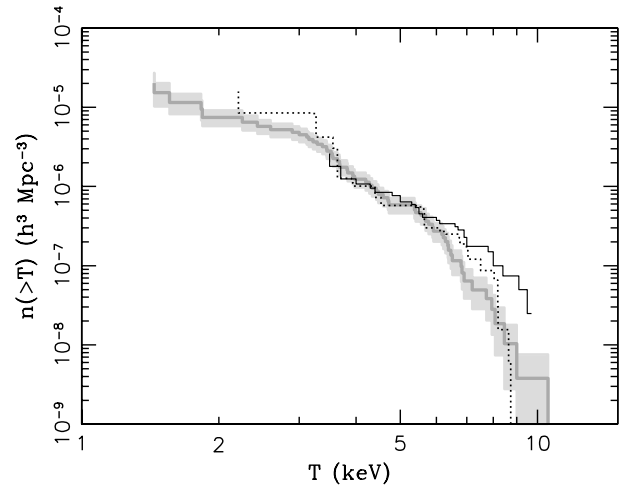,width=8.5cm}
%\caption{The cumulative temperature function for the case of a flat universe
%($\Omega_{\rm m,0}=0.2, \Omega_{\Lambda,0}=0.8$) is shown with open circles.
%The errors are estimated by 
%$\sqrt{\sum_{T_{\rm i}<T} \left( 1/V_{\rm max}(T_{\rm i}) \right)^2}$.
%Previous results from Markevitch (\cite{Markevitch}) and Henry (\cite{Henry})
%are also shown by the bold and thin lines, respectively.}
\caption{The cumulative temperature function for the case of a flat universe
($\Omega_{\rm m,0}=0.2, \Omega_{\Lambda,0}=0.8$) is shown in grey solid line
with its 68\% error band.
The errors are estimated by 
$\sqrt{\sum_{T_{\rm i}<T} \left( 1/V_{\rm max}(T_{\rm i}) \right)^2}$.
Previous results from Markevitch (\cite{Markevitch}) and Henry (\cite{Henry})
are also shown by the solid and dotted black lines, respectively.}
\label{INT_XTF}
\end{figure}

We have derived a cumulative XTF, which is the number density
of clusters that are hotter than a certain temperature, by
\begin{equation}
n(T) = \sum_{T<T_{\rm i}} \frac{1}{V_{\rm max}(T_{\rm i})}\ .
\label{eq:IntXTF}
\end{equation}
The resulting cumulative XTF is illustrated in Fig.~\ref{INT_XTF}
together with other recent results by Henry (\cite{Henry}) and 
Markevitch (\cite{Markevitch}).
Our result shows agreement with Henry's XTF constructed 
from his sample of 25 clusters.
The individual temperatures of the 25 clusters were evaluated 
by averaging the previous temperature measurements with {\it ASCA}
and other missions based on single-temperature fits (see Henry \cite{Henry}),
which agree with the hot component temperatures obtained in this work
on average within 1\%.
Above 6~keV, there is apparent disagreement between our result and
the result by Markevitch (\cite{Markevitch}) showing a higher amplitude.
This is due to a larger number of clusters in the highest temperature bins
in the Markevitch sample,
despite the fact that smaller $V_{\rm max}$ values apply to
his sample.
(Markevitch selected his sample with the same flux limit as ours
but also limited the redshift range to 0.04 $\leq$ $z$ $\leq$ 0.09).
The larger number of very hot clusters in Markevitch (\cite{Markevitch}) is
due to higher temperature estimates compared to our values
for temperatures above $\sim6$~keV (Fig.~\ref{T-T}b).

Blanchard et al. (\cite{Blanchard}),
using a sample and temperatures mainly from the XBACS
(Ebeling et al. \cite{XBACs})
and the {\it ASCA} measurement by Markevitch et al. (\cite{MarkevitchTmap}),
respectively,
obtained the XTF, which virtually perfectly agrees with that
by Markevitch (\cite{Markevitch}).
Pierpaoli et al. (\cite{Pierpaoli}), who have made another attempt to derive
the XTF, used the same sample as Markevitch (\cite{Markevitch}) 
but the temperatures given by White (\cite{White00}).
The Pierpaoli cumulative XTF has the highest amplitude among other works.
This is because of the systematically higher temperatures by 
White (\cite{White00}) (see Fig.~\ref{T-T}).

%=====4=====4=====4=====4=====4=====4=====4=====4=====4=====
\section{Constraint on cosmological parameters}
%=====4=====4=====4=====4=====4=====4=====4=====4=====4=====
%-- Sigma8 and Omega0 --\\
In the following, using the flux-limited complete sample of 61 clusters
($T_{\rm h} \geq 1.4$) and theoretical XTF models,
we obtain constraints on cosmological parameters.
The XTF at the present epoch is most sensitive to the amplitude of
density fluctuations on a scale of $8 h^{-1}$ Mpc, $\sigma_8$,
and least sensitive to the cosmological constant, $\Lambda$.
We therefore concentrate on two different $\Omega_{\Lambda,0}$ model
families including open ($\Omega_{\rm m,0}<1$, $\Omega_{\Lambda,0}=0$) 
and flat ($\Omega_{\rm m,0} + \Omega_{\Lambda,0} = 1$) universes.
Recent measurements of high-redshift SN Ia indicate a positive cosmological
constant (Perlmutter et al. \cite{Perlmutter}; Riess et al. \cite{Riess}).
Moreover, the cosmic microwave background (CMB) temperature-fluctuation 
map recently derived with the Boomerang experiment strongly supports
a flat universe model (de Bernardis et al. \cite{deBernardis}).
Therefore, a flat universe model with a positive cosmological constant
is currently the most plausible description of our universe.
For the purpose of comparison with previous works, 
we also consider the case of an open universe.

For the theoretical prediction of the XTF,
we employ two different models,
a conventional Press-Schechter mass function (Sect. 4.1)
and the Formation-Epoch model by Kitayama \& Suto (\cite{KS})
that is a modification of the former theory (Sect. 4.2).
The model XTF depends also on the Hubble constant,
which is here assumed to be 71 km/s/Mpc (i.e. $h=0.71$)
based on the latest results of the HST Hubble constant project
(see Mould et al. \cite{Mould} and references therein).

%----4.1---4.1---4.1---4.1---4.1---4.1---4.1---4.1---4.1----
\subsection{Press-Schechter model}
%----4.1---4.1---4.1---4.1---4.1---4.1---4.1---4.1---4.1----
%-- model explanation --\\
According to the Press-Schechter formalism,
the number density of clusters with a mass in the range of $M$ to $M+dM$
that have collapsed before a redshift $z$ is given as
\begin{eqnarray}
n_{ps}(M,z) dM = \hspace{5cm} & & \nonumber\\
	\sqrt{\frac{2}{\pi}}\frac{\rho_0}{M}
	\frac{\delta_c(z)}{\sigma^{2}(M)}
	\left| \frac{d\sigma (M)}{dM} \right| 
	\exp{ \left( -\frac{\delta^2_c(z)}{2\sigma^{2}(M)} \right) }
	dM \ , & &
\label{eq:PS}
\end{eqnarray}
where $\rho_0$ is the mean density of the universe at present,
$\delta_c$ is the linear overdensity of a system collapsed at a redshift
$z$ and evaluated at present
(for which we use the formula given in Kitayama \& Suto \cite{KS}),
and $\sigma^2$ is the variance of mass extracted from the density fluctuation
field within a spherical region that includes a mass $M$ on average.
For the mass variance, we employed an analytical formula derived
by Kitayama \& Suto (\cite{KS}) based on the CDM power spectrum,
which is given as
\begin{equation}
\sigma \propto (1+2.208m^{p} - 0.7669m^{2p} + 0.7949m^{3p})^{2/(9p)},
\label{eq:massvariance}
\end{equation}
where $p = 0.0873$, 
$m \equiv M(\Gamma h)^3/(\Omega_{\rm m,0} h^2)/10^{12}$.
$\Gamma$ is the shape parameter and is given as
\begin{equation}
\Gamma = \Omega_{\rm m,0} h (T_0/2.7 {\rm K})^{-2}
	\exp [ - \Omega_{\rm B} ( 1 + \sqrt{2h} \Omega_{\rm m,0}^{-1} ) ],
\label{eq:shapeparameter}
\end{equation}
where $T_0$ and $\Omega_{\rm B}$ are 
the temperature of the cosmic microwave background
and the baryon density, 
which are assumed to be 2.726~K (Mather et al. \cite{Mather})
and $\Omega_{\rm B} = 0.0193 h^{-2}$ (Burles \& Tytler \cite{BT}),
respectively.

%-- M-T relation --\\
The mass function given above is converted to the XTF by means of
a mass-temperature relation.
We assume that the ICM in a cluster is described by an isothermal
$\beta$ model (e.g. Forman \& Jones \cite{FormanJones}),
in which the density profile is given as
$ n_{\rm gas} = n_{\rm gas,0} [ 1 + (R/R_c)^{2} ]^{-3\beta/2} $,
and that hydrostatic equilibrium is achieved.
The virial mass is defined as
\begin{equation}
M_{\rm v} \equiv \frac{4}{3} \pi R_{\rm v}^3 \rho_0 (1+z)^3 \Delta 
 = \frac{3\beta kTR_{\rm v}}{\mu m_{\rm p} G}
	\frac{(R_{\rm v}/R_{\rm c})^2}{1+(R_{\rm v}/R_{\rm c})^2},
\label{eq:XrayMass}
\end{equation}
where $T$ is the temperature of ICM,
$R_{\rm v}$ is the virial radius,
$R_{\rm c}$ is the core radius,
$\mu$ is the mean molecular weight of the ICM (set to 0.6),
$m_p$ is the proton mass, $G$ is the gravitational constant,
and $\Delta$ is the ratio between a cluster mean density
and the mean density of the universe at the cluster-formation redshift $z$.
For $\Delta$ we use the formula given in Kitayama \& Suto (\cite{KS}).
Since $R_{\rm v}/R_{\rm c} >> 1$,
the virial mass is approximately given as,
\begin{equation}
M_{\rm v} = \frac{9}{2}(\pi \rho_0 \Delta)^{-1/2}
	\left( \frac{\beta}{\mu m_p G} \right)^{3/2}
	\left(\frac{kT}{1+z}\right)^{3/2}\ .
\label{eq:MT}
\end{equation}
Note that when $\beta/\beta_{\rm spec} = 2/3$,
($\beta_{\rm spec} = \frac{\mu m_p \sigma^2}{kT}$),
this equation becomes identical to that derived from a modified singular 
isothermal sphere by Bryan \& Norman (\cite{BN}).
Although an additional modification to the $M-T$ relation is often applied
by multiplying a factor accounting for departures from virial equilibrium
(e.g. Henry \cite{Henry}), we did not introduce this factor here.
The observed values of $\beta$ are distributed in a relatively
narrow range around $\sim 2/3$ and
may show a weakly positive correlation with temperatures
(Arnaud \& Evrard \cite{AE}; Vikhlinin et al. \cite{VFJ};
Horner et al. \cite{HMS}).
Vikhlinin et al. measured $\beta$ values systematically
with {\it ROSAT} PSPC data excluding a central region inside 30\% of 
the virial radius and found a correlation of $\beta$ with X-ray temperature,
which can be fitted with the linear function
$\beta = 0.017\ T({\rm keV}) + 0.621 $ (Henry \cite{Henry}).
We employ this relation and substitute it into Eq. \ref{eq:MT}.
The effect of using different $M-T$ relations
will be studied in Sect. 5.

%-- Recent formation approximation--\\
We further assume that a cluster collapsed just before it is observed.
This recent formation approximation, which has been extensively
applied in previous conventional modelings, is relaxed 
in the Formation-Epoch model described in the next section.
Combining all the formulae given above,
the model XTF at a given redshift is given by
\begin{equation}
\phi_{\rm ps}(T,z)\ dT 
	= n_{\rm ps}(M_{\rm v}(T,z),z)\ \frac{dM_{\rm v}}{dT}(T,z)\ dT \ .
\label{eq:PSXTF}
\end{equation}
Hereafter we refer to this model XTF as the PS model.

%----4.2---4.2---4.2---4.2---4.2---4.2---4.2---4.2---4.2----
\subsection{Formation-Epoch model}
%----4.2---4.2---4.2---4.2---4.2---4.2---4.2---4.2---4.2----
%-- Limitation with the conventional XTF and introduction to KS model --\\
In the model XTF given in the previous section,
the formation redshift, $z_f$, of an observed cluster
is assumed to be the same as the observed redshift
(i.e. recent formation approximation).
However, this assumption is not reasonable in particular
in a low density universe.
Kitayama \& Suto (\cite{KS}) constructed a model XTF 
taking into account the distribution of the dark-matter-halo formation epoch
as well as the evolution of temperature (hereafter FE model).
They showed that the XTF predicted by the FE model can be significantly 
different from the PS model described in Sect. 4.1,
in particular when the temperature evolves after the collapse.

%-- FE model description --\\
The FE model is given as,
\begin{eqnarray}
\phi_{\rm FE}(T_{\rm obs},z) dT_{\rm obs}
 = \int^{\infty}_{z} dz_{\rm f} 
\left[ 2 \frac{\partial p}{\partial z_{\rm f}}(z_{\rm f},2M_{\rm v}(T,z_{\rm f}),z)\right.
 \hspace{0cm}  & & \nonumber\\
\left. n_{\rm ps}(2M_{\rm v}(T,z_{\rm f}),z) \frac{1}{\kappa (z_{\rm f},z)}
 \frac{dM_{\rm v}}{dT}(T,z_{\rm f})\right] dT_{\rm obs} \ , & &
\label{eq:FEXTF}
\end{eqnarray}
where $n_{\rm ps}$ and $M_{\rm v}$ are given in Eqs. \ref{eq:PS}
and \ref{eq:MT}, respectively, and
$\partial p / \partial z_{\rm f}$ given by Lacey \& Cole (\cite{LC})
is a differential distribution function of halo formation epochs
of a cluster that has a mass of $2M_{\rm v}$ at the observed redshift $z$.
At the collapse redshift $z_{\rm f}$, the cluster is assumed to be formed
with a mass of $M_{\rm v}$, increasing in mass after its formation.
$\kappa(z_{\rm f},z) = \left( \frac{1+z_{\rm f}}{1+z} \right)^s$ gives
the evolution in temperature via
$T_{\rm obs}(z_{\rm f},z) = \kappa (z_{\rm f},z) T(z_{\rm f})$,
where $T(z_{\rm f})$ is the virial temperature
at the collapse redshift $z_{\rm f}$,
while $T_{\rm obs}(z_{\rm f},z)$ is the observed temperature
at the redshift $z$.
$s=0$ corresponds to no temperature evolution,
while $s>0$ and $s<0$ indicate positive and negative temperature evolution
after the collapse, respectively.

%----4.3---4.3---4.3---4.3---4.3---4.3---4.3---4.3---4.3----
\subsection{Fitting method}
%----4.3---4.3---4.3---4.3---4.3---4.3---4.3---4.3---4.3----
%-- fitting --\\
Instead of fitting the model XTF to the XTF derived in Sect. 3,
we use the predicted XTF and the $L-T$ relation
to calculate the expected luminosity and temperature distribution
for the given survey selection function which is given 
in a form of $V_{\rm max}$,
and perform a fit to the observed luminosity-temperature distribution
of the 61 clusters selected in Sect. 2.4.
In this analysis,
the intrinsic luminosity distribution for a given temperature,
which in other words is the conditional luminosity function 
at a given temperature,
is assumed to follow a Gaussian function in logarithmic scale
with a constant standard deviation
and a mean luminosity given by a power-law function of the temperature.
This assumption can be justified by observational results as shown
in Sect. 3.1.
In order to simplify a model calculation,
we assume that the XTF does not change within the search volume
of our flux limit
and use a model XTF at the median redshift of our cluster sample 
at z=0.046 as a representative value.
Systematic errors on the final results caused from this assumption
are found to be negligibly small compared with statistical errors.
Incorporated by another Gaussian that represents temperature
measurement errors,
the expected cluster number density 
in a unit logarithmic temperature and a unit logarithmic luminosity
is given as
\begin{eqnarray}
N(L,T)\ d \log L\ d\log T = \hspace{4.6cm} & & \nonumber\\
 \int^{\infty}_{-\infty} dT' \left[ \frac{\phi(T',z=0.046)}{(h=0.71)^{3}} 
 v_{\rm max}(L,T') \right. \hspace{2cm} & & \nonumber\\
 \frac{1}{\sqrt{2\pi \sigma_{\log L}^2}}
 \exp \left\{ -\frac{(\log AT^{\alpha}-\log L)^2}{2\sigma_{\log L}^2}\right\}
 \hspace{2cm} & & \nonumber\\
  \left. \frac{1}{\sqrt{2\pi \sigma_{\rm T}^2}}
  \exp \left\{ -\frac{(T'-T)^2}{2\sigma_{\rm T}^2}\right\}
  \frac{dT}{d\log T} \right] d\log L\ d\log T,\ \ \ & &
\label{eq:FF}
\end{eqnarray}
where $\phi$ is calculated for $h=0.71$ and normalized with the $h^{-3}$
factor to a case of the $h=1$ universe.
$v_{\rm max}(L,T)$ is a maximum comoving volume calculated for $h=1$,
in which a cluster having a temperature of $T$
and a luminosity of $L$ could have been detected
under our sample selection criteria,
for which the calculation method is given by Eq. \ref{eq:modvmax}.
The temperature measurement errors are
$\sigma_{\rm T}$ = $0.051 + 0.0014\ T^{2.585}$
from our {\it ASCA} results.
Free parameters to a fit are;
$\sigma_8$ and $\Omega_{\rm m,0}$ specifying $\phi$;
$A$ and $\alpha$ specifying the power-law function of the $L-T$ relation
that gives the mean of the luminosity;
and $\sigma_{\log L}$ that is the constant standard deviation
of the luminosity from the $L-T$ relation in logarithmic scale.
Although $v_{\rm max}$ varies with the assumed model universe,
i.e. with the values of $\Omega_{\rm m,0}$ as well as $\sigma_8$
which are the fitting parameters,
we used a specific $v_{\rm max}$ value calculated with
($\Omega_{\rm m,0}, \Omega_{\Lambda,0}$) = (0.2, 0), or (0.2,0.8),
in the case of an open or flat universe, respectively.
Since all the clusters in our sample are nearby,
$v_{\rm max}$ is virtually constant for different $\Omega_{\rm m,0}$ values,
using the fixed $v_{\rm max}$ does not introduce any considerable
systematic error on the final constraints on $\Omega_{\rm m,0}$
and $\Omega_{\Lambda,0}$.
The model cluster distribution was then fitted to the observed
temperature and luminosity distribution.
Assuming a Poisson distribution of cluster number counts 
at a given luminosity and temperature,
we defined a logarithmic likelihood function as
\begin{equation}
\ln {\cal L} = \sum_{\rm i} \ln N(T_{\rm i},L_{\rm i}) 
- \int N(T,L)\ d\log L\ d\log T \ ,
\label{eq:LH}
\end{equation}
where $i$ denotes each sample cluster and the integration
is performed over the $\log L$(ergs s$^{-1}$) range from 41.5 to 45.5
and the $\log T$(keV) range from 0.14 to 1.05
(corresponding to T = 1.4 -- 11.2~keV).
By maximizing the likelihood function,
we derived the best-fit model XTF as well as the $L-T$ relation.

There are two reasons for setting the low temperature cut-off at 1.4~keV.
First, as discussed in Sect. 2.3,
the $L-T$ relation breaks down for elliptical galaxies which have
a lower gas mass fraction, and actually gets steeper below $\sim1$~keV
as shown by e.g. Xue \& Wu (\cite{XW}).
Thus the simple power-law modeling for the $L-T$ relation is not appropriate
for the whole temperature range.
The second reason is based on the more essential problem concerned
with the comparison between the theoretical XTFs and the observation.
The systems at the low temperature end are mostly single elliptical galaxies
and some of them may not be isolated but surrounded by larger scale dark
matter halos.
As has been shown by e.g. Matsushita (\cite{MatsushitaPhD}),
X-ray brightness profiles of elliptical galaxies have large variety
suggesting a significant variation in total mass for a given temperature,
i.e. a considerable scatter in the $M-T$ relation would be expected.
Therefore, a cluster abundance at the low temperature end probably
does not directly correspond to an abundance of dark matter halos
at a certain mass scale.

%----4.4---4.4---4.4---4.4---4.4---4.4---4.4---4.4---4.4----
\subsection{Results}
\begin{figure}
\centerline{
\psfig{file=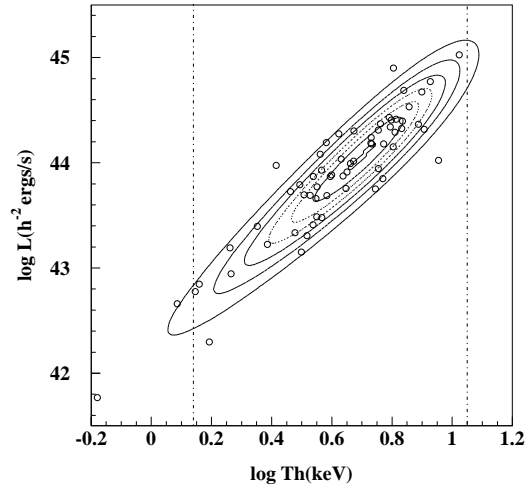,width=8.0cm,angle=0,clip=}}
\caption{The best-fit model function in the luminosity-temperature space
for the case of the flat universe is shown with contours. 
The 63 clusters from the flux-limited complete sample
are overlayed with open circles. 
The vertical dotted lines indicate the temperature range used for the fitting.
Therefore, 61 clusters are used to make constraints.}
\label{lL-lT}
\end{figure}

\begin{figure}
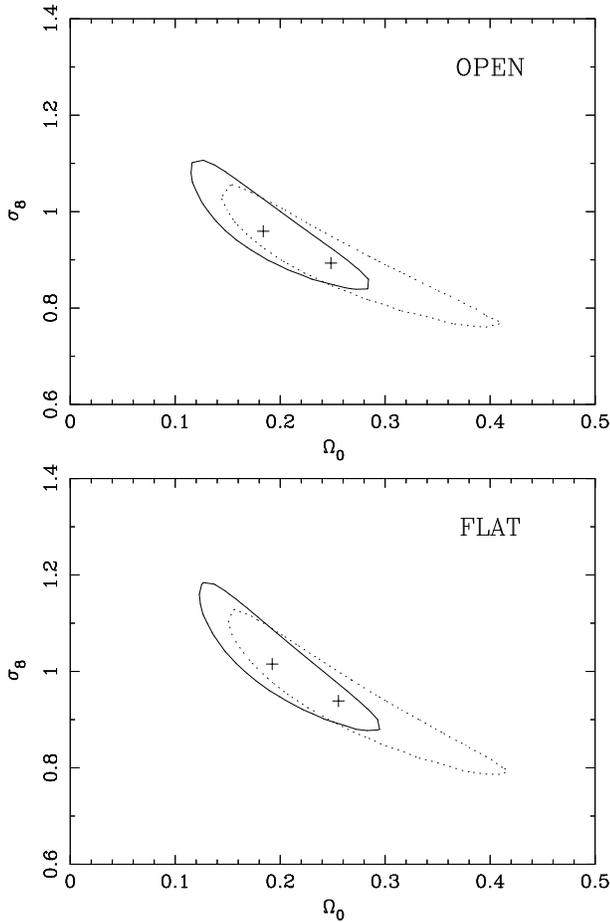

\centerline{
\psfig{file=h2920f8a.ps,width=8.0cm,angle=-90,clip=}}
\centerline{
\psfig{file=h2920f8b.ps,width=8.0cm,angle=-90,clip=}}
\caption{Constraints on $\Omega_{\rm m,0}$ and $\sigma_8$ with the PS model
in case of the open (upper panel) and flat (lower panel) universe.
Bold contours are results including all sample clusters hotter than 1.4~keV,
while dotted contours are derived when only clusters hotter than 3.0~keV
are used.}
\label{PS_CONT}
\end{figure}

%----4.4---4.4---4.4---4.4---4.4---4.4---4.4---4.4---4.4----
We first present the results from the PS model fitting.
The best-fit values and errors of the fitting parameters
are summarized in Table 3.
The best-fit distribution function in the log$L-$log$T$ space
in the flat universe case is illustrated in Fig.~\ref{lL-lT}.
The corresponding model XTF is also illustrated in Fig.~\ref{XTF+SIMPLE}.
The constraint in $\Omega_{\rm m,0} - \sigma_8$ space is shown in
Fig.~\ref{PS_CONT} for each open and flat universe case.
The contours represent the 90\% confidence range for two parameters
of interest, and the constraint is notably tight.

In order to demonstrate an improvement on the constraint
by the widest temperature range and the largest sample ever achieved,
we also performed the fitting with the PS model
only using 51 sample clusters above 3~keV.
As shown in Fig.~\ref{PS_CONT}, the improvement is significant,
in particular, for $\Omega_{\rm m,0}$.
Since $\Omega_{\rm m,0}$ is tightly related to the overall shape of
the mass variance, i.e. the power spectrum,
a larger mass range to measure the mass fluctuation amplitude
is essential.

The ``true'' $L-T$ relation employed in Sect. 2.3 and Sect. 3.1
was estimated by this fitting.
Fixing the $\alpha$ value as obtained in Sect. 2.3,
to 2.47 and 2.50 for the open and flat universe case, respectively,
we performed the PS model fitting and derived
$\log A$=42.15, $\sigma=0.24$ for the open universe case,
and $\log A$=42.14, $\sigma=0.24$ for the flat universe case,
which are given in Eqs. \ref{revLTopen} and \ref{revLTflat}
and are applied in Eq. \ref{eq:VmaxT}.

\begin{figure}
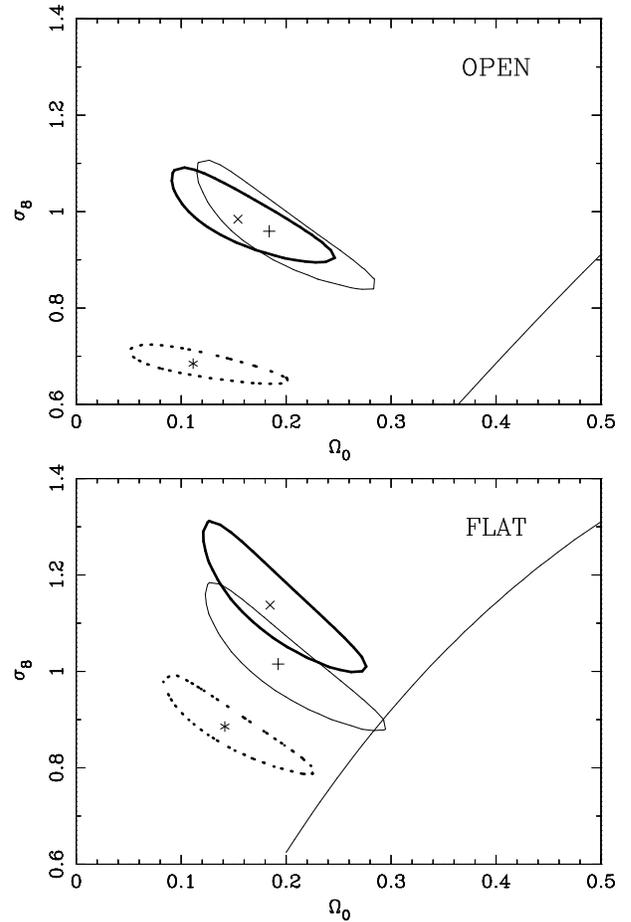

\centerline{
\psfig{file=h2920f9a.ps,width=8.0cm,angle=-90,clip=}}
\centerline{
\psfig{file=h2920f9b.ps,width=8.0cm,angle=-90,clip=}}
\caption{Constraints on $\Omega_{\rm m,0}$ and $\sigma_8$ in case of 
the open (upper panel) and flat (lower panel) universe.
Thin closed contours show results with the PS model.
Bold dotted and solid closed contours are derived with
the $s=1$ and $s=0$ FE models, 
with and without a temperature evolution effect, respectively.
The line shows the constraints from the {\it COBE} normalization
by Bunn \& While (\cite{BW}) for $h=0.71$.}
\label{FE_CONT}
\end{figure}

The FE model gives somewhat different results.
In addition to $\Omega_{\rm m,0}$ and $\sigma_8$,
the FE model introduces another parameter, $s$, that models
a temperature evolution, which can be a free parameter in a fit.
However, following Kitayama \& Suto (\cite{KS}), in the following analysis,
we rather fixed $s$ at $s=0$ and $s=1$ such that they represent no evolution
and a maximum possible positive evolution of the temperature, respectively.
Since there is no observational evidence for the evolution of the temperature,
e.g. no redshift dependency in the $L-T$ relation is found
(Mushotzky \& Scharf \cite{MushotzyScharf};
Donahue et al. \cite{Donahue99}; Della Ceca et al. \cite{DellaCeca00}),
the no evolution model may be more plausible.
However, 
the effect of the temperature evolution is still interesting to study here.
The fitting results are summarized in Table 3,
and the constraints on $\Omega_{\rm m,0}$ and $\sigma_8$ are illustrated
in Fig.~\ref{FE_CONT}.
The FE model in the no temperature evolution case ($s=0$)
shows similar results with those of the PS model,
since the XTF predicted by the $s=0$ FE model
is very close to that of the PS model
(see Fig.8 in Kitayama \& Suto \cite{KS}).
On the other hand,
in the maximum temperature evolution cases ($s=1$),
the FE model gives significantly smaller $\sigma_8$ values
than those with the PS model.

\begin{table*}
\begin{center}
\caption{Fitting Results}
\begin{tabular}{lccccc}
\hline %------------------------------------------------
\hline %------------------------------------------------
Model XTF & $\Omega_{\rm m,0}$ & $\sigma_8$ & A & $\alpha$ & $\sigma_{\log L}$ \\
\hline %------------------------------------------------
          PS(Open) & 0.18$_{-0.05}^{+0.08}$ & 0.96$_{-0.09}^{+0.11}$ & 42.15$_{-0.20}^{+0.17}$ & 2.47$_{-0.24}^{+0.26}$ & 0.24$_{-0.04}^{+0.04}$ \\
 PS(Open,$T>3$keV) & 0.25$_{-0.09}^{+0.12}$ & 0.89$_{-0.11}^{+0.13}$ & 42.19$_{-0.28}^{+0.27}$ & 2.42$_{-0.37}^{+0.38}$ & 0.22$_{-0.03}^{+0.05}$ \\
      FE(Open,s=0) & 0.15$_{-0.05}^{+0.07}$ & 0.98$_{-0.07}^{+0.08}$ & 42.14$_{-0.20}^{+0.17}$ & 2.49$_{-0.25}^{+0.26}$ & 0.24$_{-0.04}^{+0.04}$ \\
      FE(Open,s=1) & 0.11$_{-0.05}^{+0.07}$ & 0.68$_{-0.03}^{+0.04}$ & 42.12$_{-0.20}^{+0.18}$ & 2.51$_{-0.25}^{+0.27}$ & 0.24$_{-0.04}^{+0.04}$ \\
          PS(Flat) & 0.19$_{-0.05}^{+0.08}$ & 1.02$_{-0.11}^{+0.12}$ & 42.15$_{-0.20}^{+0.17}$ & 2.49$_{-0.25}^{+0.26}$ & 0.24$_{-0.03}^{+0.04}$ \\
 PS(Flat,$T>3$keV) & 0.26$_{-0.09}^{+0.12}$ & 0.94$_{-0.13}^{+0.14}$ & 42.19$_{-0.29}^{+0.27}$ & 2.44$_{-0.38}^{+0.39}$ & 0.23$_{-0.04}^{+0.04}$ \\
      FE(Flat,s=0) & 0.18$_{-0.05}^{+0.07}$ & 1.14$_{-0.11}^{+0.13}$ & 42.15$_{-0.20}^{+0.17}$ & 2.49$_{-0.25}^{+0.26}$ & 0.24$_{-0.03}^{+0.04}$ \\
      FE(Flat,s=1) & 0.14$_{-0.05}^{+0.06}$ & 0.89$_{-0.08}^{+0.08}$ & 42.14$_{-0.21}^{+0.17}$ & 2.51$_{-0.25}^{+0.26}$ & 0.24$_{-0.03}^{+0.04}$ \\
\hline %------------------------------------------------
\end{tabular}
\end{center}
\begin{itemize}
\setlength{\itemsep}{2mm}
\item The error ranges corresponding to 90\% confidence for
one parameter of interest.
\end{itemize}
\label{FitRes}
\end{table*}

%=====5=====5=====5=====5=====5=====5=====5=====5=====5=====
\section{Discussion}
%=====5=====5=====5=====5=====5=====5=====5=====5=====5=====
%----5.1---5.1---5.1---5.1---5.1---5.1---5.1---5.1---5.1----
\subsection{Uncertainty in the $M-T$ relation}
%----5.1---5.1---5.1---5.1---5.1---5.1---5.1---5.1---5.1----
The $M-T$ relation used in our analysis may be subject to uncertainties
introducing a significant systematic error in the final results 
on the cosmological parameters.
Theoretical arguments yield a scaling law, that is
$M \propto T^{3/2}$ (e.g. Kaiser \cite{Kaiser}).
A number of numerical simulations support this scaling law,
but different normalizations are derived from different simulations.
Henry (\cite{Henry}) summarized such differences in the normalization
of the $M-T$ relation derived from different hydrodynamical simulations.
A cluster mass may vary from 62\% to 119\% of the mean value.

The $M-T$ relations derived observationally from X-ray mass measurements
give somewhat different results.
Horner et al. (\cite{HMS}) derived an $M-T$ relation which has
a similar exponent as the scaling law but a lower normalization than
the numerical simulations.
On the other hand, Nevalainen et al. (\cite{NMF})
and Finoguenov et al. (\cite{FRB}) obtained steeper relations
of $T^{1.79}$ and $T^{1.78}$, respectively,
and lower normalizations.

\begin{figure}
\centerline{
\psfig{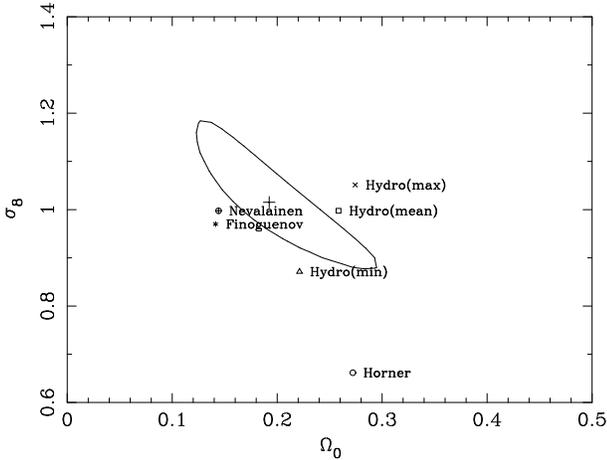}}
\caption{The best-fit values of $\Omega_{\rm m,0}$ and $\sigma_8$
derived with the PS model in the case of the flat universe,
for which different $M-T$ relations are applied.
A 90\% confidence contour derived in Sect. 4 is also overlayed.}
\label{difM-T}
\end{figure}

In order to investigate the effects of the different $M-T$ relations,
we repeated the fitting performed in Sect. 4 with the PS model
for the flat universe case.
The six $M-T$ relations we have examined are summarized in Table 4.
Each $M-T$ relation can be expressed with Eq. \ref{eq:MT},
replacing the $\beta$ value with the formula given in Table 4.
In Fig.~\ref{difM-T}, the best-fit parameters derived with the different
$M-T$ relations are compared with the result presented in Sect. 4.4.
The effect of using different $M-T$ relations is significant
and the systematic errors introduced are comparable to, or even larger than,
the current statistical errors.
As seen in Fig.~\ref{difM-T}, a change in the index of the $M-T$ relation
moves the best-fit point along the elongation direction
of the confidence contour, while an amplitude change moves the best-fit
point along the perpendicular direction.
Taking into account these systematics from the different $M-T$ relations,
the constraints on $\Omega_{\rm m,0}$ and $\sigma_8$ are revised
and summarized in Table 5.

With the FE model, where the recent formation approximation is not valid,
it is not appropriate to use the empirical $M-T$ relation obtained based 
on the observations of nearby clusters.
This is because 
the current $M-T$ relation should be a result of accumulation
of different $M-T$ relations from different formation redshifts which
depend on the cluster masses.
Therefore, the $M-T$ relations obtained from the Hydrodynamic simulations
may be more appropriately applied with the FE model.

\begin{table}
\begin{center}
\caption{M-T relations}
\begin{tabular}{cc}
\hline %------------------------------------------------
\hline %------------------------------------------------
      Source                      & $\beta$ \\
\hline %------------------------------------------------
Hydrodynamic simulation (mean)    & $\frac{2}{3} \times 1.21 $ \\
Hydrodynamic simulation (maximum) & $\frac{2}{3} \times 1.36 $ \\
Hydrodynamic simulation (minimum) & $\frac{2}{3} \times 0.88 $ \\
Horner et al.                      & 0.496 $T^{-0.133}$ \\
Nevalainen et al.                 & 0.437 $T^{0.193}$ \\
Finoguenov et al.                 & 0.415 $T^{0.187}$ \\
\hline %------------------------------------------------
\end{tabular}
\end{center}
\label{tab:MT}
\end{table}

\begin{table}
\begin{center}
\caption{Results with different M-T relations.
The parameter ranges correspond to 90\% statistical error
for one parameter of interest
plus systematic errors with different $M-T$ relations applied.}
\begin{tabular}{lcc}
\hline %------------------------------------------------
\hline %------------------------------------------------
Model XTF & $\Omega_{\rm m,0}$ & $\sigma_8$ \\
\hline %------------------------------------------------
          PS(Open) & 0.09 -- 0.37 & 0.55 -- 1.13 \\
      FE(Open,s=0) & 0.07 -- 0.32 & 0.61 -- 1.15 \\
      FE(Open,s=1) & 0.03 -- 0.23 & 0.52 -- 0.80 \\
          PS(Flat) & 0.10 -- 0.38 & 0.57 -- 1.19 \\
      FE(Flat,s=0) & 0.10 -- 0.35 & 0.65 -- 1.33 \\
      FE(Flat,s=1) & 0.06 -- 0.18 & 0.66 -- 0.97 \\
\hline %------------------------------------------------
\end{tabular}
\end{center}
\label{ResMT}
\end{table}

%----5.2---5.2---5.2---5.2---5.2---5.2---5.2---5.2---5.2----
\subsection{Comparison with other measurements}
%----5.2---5.2---5.2---5.2---5.2---5.2---5.2---5.2---5.2----
We compared our constraints on the cosmological parameters
with previous works.
Bahcall (\cite{Bahcall00}) summarized the cluster constraints
on $\Omega_{\rm m,0}$ and $\sigma_8$ as $0.2\pm0.1$ and
$1.2\pm0.2$ (68\% confidence level), respectively.
These values are consistent with our results.

The power spectrum of matter distribution in the present universe
has been measured from the spatial distribution of galaxies 
and galaxy clusters,
and found to be generally well reproduced by the CDM power spectrum
(Peacock \& West \cite{PW}; Einasto et al. \cite{Einasto93};
Jing \& Valadarnini \cite{JV}; Einasto et al. \cite{Einasto97};
Retzlaff et al. \cite{Retzlaff98};
Tadros et al. \cite{TED}; Miller \& Batuski \cite{MB}).
An important feature of the power spectra is a turnover,
where the scale is closely linked to the horizon scale at matter-radiation
equality.
Most recently Schuecker et al. (\cite{REFLEXII}),
using a sample of 452 X-ray selected clusters of galaxies,
determined the location of the maximum of the power spectrum
in the range of $0.022 < k_{\rm max} < 0.030$ $h$ Mpc$^{-1}$.
According to the CDM power spectrum, $k_{\rm max}$ is well approximated by
\begin{equation}
k_{\rm max} = 0.114\ \Gamma\ h\ {\rm Mpc}^{-1} \ ,
\label{eq:k}
\end{equation}
where $\Gamma$, given in Eq. \ref{eq:shapeparameter},
is a function of $\Omega_{\rm m,0}$.
Therefore one finds $0.33 < \Omega_{\rm m,0} < 0.44$.
The constraints on $\Omega_{\rm m,0}$ we have derived in Sect. 5.1
with the PS model yield $0.005 < k_{\rm max} < 0.027 $ $h$ Mpc$^{-1}$,
which overlaps with the result by Schuecker et al.
Our results with the FE model give 
$4\times10^{-4} < k_{\rm max} < 0.022 $ $h$ Mpc$^{-1}$
in the case of the open universe,
and $2\times10^{-3} < k_{\rm max} < 0.025 $ $h$ Mpc$^{-1}$
in the case of the flat universe,
both of which are also consistent with the result by Schuecker et al.

A measurement of the matter density fluctuation at the largest
spatial scale was performed using the {\it COBE} DMR experiment
(e.g. Bennett et al. \cite{Bennett}).
Bunn \& White (\cite{BW}), using the CDM power spectrum model,
extrapolated the {\it COBE} results to cluster scales
and gave constraints in $\Omega_{\rm m,0}$-$\sigma_8$ space,
which are shown in Fig.~\ref{FE_CONT}.
In the case of the open universe,
our results with any of the model XTFs
do not overlap the {\it COBE} constraints,
even if the effects of the uncertainty in the $M-T$ relation
are taken into account.
On the other hand, in the case of the flat universe,
our results with the effects of the uncertainty in the $M-T$ relation,
are consistent with the {\it COBE} constraints
regardless of the choice of the model XTF.

%=====6=====6=====6=====6=====6=====6=====6=====6=====6=====
\section{Summary and Conclusions}
%=====6=====6=====6=====6=====6=====6=====6=====6=====6=====
The most important result concerning the XTF derived here
is that it covers a large enough range of temperatures
to reveal the shape of the function beyond a simple power law
representation.
As expected from theoretical considerations,
there should be a very sharp cut-off at the high mass end
and the function should turn into a shallower slope
at the low mass end as shown by our observations.
We have demonstrated that these constraints on the actual slope
of the XTF allow us to derive independent constraints on the two parameters
which are most important in cosmic structure formation models:
$\Omega_{\rm m,0}$ and $\sigma_8$.
Leaving out only 16\% of the sample clusters from the low temperature end
almost doubles the uncertainty for these independent constraints.
Therefore, while these independent constraints are an important achievement
of the present work on one hand,
this demonstration also shows that 
these constraints are sensitive to any systematic error
introduced into the XTF.

For this reason we have studied the stability of our results to
variations introduced to the observed data and also variations
of the theoretical modeling.
We have fully incorporated the measurement errors
in fluxes and temperatures of the sample clusters.
With the typical errors of 5\% for the flux and 6-10\% for the temperature,
the biases were found to be relatively small and not to cause a drastic change
on the final results in our analysis.
On the other hand,
we have shown that there are considerable variations
and uncertainties in the theoretical modeling of the XTF,
which introduce significant systematic errors in the final constraints
on $\Omega_{\rm m,0}$ and $\sigma_8$.
Various $M-T$ relations that are employed to model the XTF
have resulted in considerably different values of $\Omega_{\rm m,0}$
and $\sigma_8$.
Another source of variation in the theoretical modeling 
that we have studied is
a recent formation approximation which is employed with
the Press-Schechter formalism to build the PS model here.
Using the FE model,
in which the distribution of cluster formation redshifts
and the evolution of the mass as well as temperature of a cluster after
the collapse are analytically formulated,
we have shown that a considerable difference could be introduced
in the derived cosmological parameters, in particular in $\sigma_8$.
This further motivates us to establish a more sophisticated model
to the XTF incorporating new X-ray observations
of the high redshift clusters with Chandra and XMM-Newton.

Finally, taking into account all the uncertainties,
we put our general constraints on $\Omega_{\rm m,0}$ and $\sigma_8$
as $\Omega_{\rm m,0}$=0.03--0.37, $\sigma_8$=0.52--1.15 in an open universe
and $\Omega_{\rm m,0}$=0.06--0.38, $\sigma_8$=0.57--1.33 in a flat universe,
respectively.
From the comparison with the {\it COBE} constraints,
a flat universe is more preferable than an open universe.

\appendix
%=====A=====A=====A=====A=====A=====A=====A=====A=====A=====
\section{Calculation of $V_{\rm max}$}
%=====A=====A=====A=====A=====A=====A=====A=====A=====A=====
Here we present our method, used in Sect. 3,
to evaluate the X-ray temperature function in more detail.
A flux limited sample has been extensively used for constructing 
the X-ray luminosity function (XLF) by many authors
(e.g. Piccinotti et al. \cite{Piccinotti};
Edge et al. \cite{Edge}; Gioia et al. \cite{Gioia};
Henry et al. \cite{Henry92}; Ebeling et al. \cite{Ebeling};
Rosati et al. \cite{Rosati}; de Grandi et al. \cite{deGrandiXLF};
Ledlow et al. \cite{Ledlow}).
Here we present the first description of how to construct
this function in the presence of measurement
uncertainties, correlation scatter and selection biases.
For a sample with a finite number of objects,
the differential XLF, i.e. the number density of clusters
having the rest frame luminosity in a range of
$L-\frac{\Delta L}{2}$ to $L+\frac{\Delta L}{2}$
defined in a comoving space, can be evaluated as
\begin{equation}
\phi(L)\Delta L = 
	\sum_{L-\frac{\Delta L}{2}\le L_{\rm i}<L+\frac{\Delta L}{2}}
	\frac{1}{v_{\rm max}(L_{\rm i},T_{\rm i})}\ ,
\label{eq:XLF}
\end{equation}
where $i$ denotes individual clusters in the luminosity range of a bin.
$v_{\rm max}(L,T)$ is a maximum search volume measured
in comoving space where a cluster having the rest frame luminosity
of $L$ and the temperature of $T$ could have been detected 
under the flux limited condition.
Introducing $z_{\rm max}$, the maximum redshift at which such a cluster
could have been detected,
and $V_{\rm com}$, a total volume in the comoving coordinate 
from us to a given redshift,
we can derive $v_{\rm max}(L,T)$ as
\begin{equation}
v_{\rm max}(L,T) = \frac{\Omega}{4\pi} V_{\rm com}(z_{\rm max})\ ,
\label{eq:vmax}
\end{equation}
where $\Omega$ is the total sky coverage in steradians,
and $V_{\rm com}$ is, in the case of a flat universe, simply given by
\begin{equation}
V_{\rm com}(z) = \frac{4\pi}{3}\left\{\frac{D_{\rm l}(z)}{1+z}\right\}^{3}\ ,
\label{eq:Vcom}
\end{equation}
where $D_{\rm l}$ is a luminosity distance,
while, in the case of an open universe,
it is given with a more complicated form (see e.g. Sandage \cite{sandage}).
$z_{\rm max}$ is given by solving the following equation
\begin{equation}
D_{\rm l} (z_{\rm max})^{2}
= \frac{L_{\rm obs}(L,T,z_{\rm max})}
  {4\pi f_{\rm obs,lim}}
\ ,
\label{eq:Dlzmax}
\end{equation}
where $f_{\rm obs,lim}$ is the flux limit defined in a certain
energy range in the observed frame,
and $L_{\rm obs}$ is the same observed-frame luminosity
that is expected if the cluster is located at $z_{\rm max}$.
In order to solve this equation with the $K$-correction,
the spectral energy distribution, i.e. the temperature, $T$, is required,
although the dependency of $z_{\rm max}$ on $T$ is generally weak.

Using $v_{\rm max}(L,T)$, the differential XTF can be given as
\begin{equation}
\phi(T)\Delta T =
	\sum_{T-\frac{\Delta T}{2}\le T_{\rm i}<T+\frac{\Delta T}{2}}
	\frac{1}{v_{\rm max}(L_{\rm i},T_{\rm i})}\ .
\label{eq:diffXTF}
\end{equation}
This estimator is used in e.g. Henry (\cite{Henry}).
However, since there is a distribution of luminosities
for a given temperature, this method introduces a scatter in a derived XTF,
unless there are large numbers of clusters in each temperature bin.
If the distribution of the luminosity for a given temperature
is known, we can obtain an average search volume by summing
$v_{\rm max}(L,T)$ in a whole range of luminosities
weighted with the probability.
Assuming that luminosities for a given temperature are distributed
like a Gaussian with a mean given by the power-law function,
$L=AT^{\alpha}$, 
and with a constant standard deviation, in logarithmic scale,
we can obtain the average search volume at a given temperature as
\begin{eqnarray}
V_{\rm max}(T) = \hspace{7cm} & & \nonumber\\
 \int^{\infty}_{-\infty} 
 \frac{v_{\rm max}(L,T)}{\sqrt{2\pi \sigma_{\log L}^2}}
 \exp\left[{-\frac{(\log AT^{\alpha} - \log L)^2}{2\sigma_{\log L}^2}}\right]
 d\log L .\ \ \ \ \ & &
\label{aeq:VmaxT}
\end{eqnarray}
We can then evaluate the temperature function as
\begin{equation}
\phi(T)\Delta T =
	\sum_{T-\frac{\Delta T}{2}\le T_{\rm i}<T+\frac{\Delta T}{2}}
	\frac{1}{V_{\rm max}(T_{\rm i})}\ .
\end{equation}
Performing Monte-Carlo simulations,
we confirmed that both estimators, the $1/v_{\rm max}(L,T)$ method
and $1/V_{\rm max}(T)$ method are consistent and unbiased,
that is they reproduce the temperature function fed to the simulations
in the limit of an infinite number of simulated clusters,
and the mean value of the temperature functions obtained 
from a number of independent measurements with a finite number of clusters
is also equal to the true value in the limit of an infinite number of
measurements. 
However, the $1/V_{\rm max}(T)$ method gives smaller variance
in a measurement of the temperature function
and therefore it is more efficient than the $1/v_{\rm max}(L,T)$ method.
This is well demonstrated in Fig.~\ref{XTF+SIMPLE}.

When we apply this analysis method to our cluster sample,
we further consider the following two points:
Our first concern is that the flux measurement may have a redshift bias
such that for clusters with the same true luminosity,
different luminosities are derived for different redshifts.
In order to study this effect we produced a sample of cluster images
at various redshifts 
as observed by the {\it ROSAT} PSPC
and determined their fluxes 
with the growth curve analysis method (B\"{o}hringer et al. 2000) 
as performed for the actual observed data of the present sample
by Reiprich \& B\"{o}hringer (\cite{RB}).
The simulations show that the resulting luminosities are kept constant
regardless of the redshift --- at least in a range that we are considering
here, i.e. $z<0.2$,
despite the overall signal to noise ratio worsening
with a dimming of the surface brightness that follows $(1+z)^{-4}$.

Secondly, the effect from the measurement errors on X-ray fluxes
is considered.
If there are considerable measurement errors in the flux measurements,
a cluster that should be in the flux-limited sample
could be found to be fainter than the flux limit
and to be missed from the sample, and vice-versa.
If the measured flux distribution is assumed to be following
a Gaussian function with the sigma value of $\sigma_{\rm f}$,
the probability that a cluster of the luminosity $L$
at the redshift of $z$ is found to be brighter
than the flux limit $f_{\rm lim}$ is given as
\begin{equation}
W(L,z,f_{\rm lim})
= \int^{\infty}_{f_{\rm lim}} \frac{1}{\sqrt{2\pi \sigma_{\rm f}^2}}
	\exp\left[{-\frac{(f-f_0)^2}{2\sigma_{\rm f}^2}}\right] df \ ,
\label{eq:W}
\end{equation}
where $f_0 = L/ 4\pi D_{\rm l}(z)^2$.
Therefore, Eq. \ref{eq:vmax} should be modified as
\begin{equation}
v_{\rm max}(L,T)
= \frac{\Omega}{4\pi} \int^{\infty}_{0} \frac{dV_{\rm com}}{dz}
 W(L,D_{\rm l},f_{\rm lim}) dz\ .
\label{eq:modvmax}
\end{equation}
Setting $\sigma_{\rm f}$ to $10^{-12}$ ergs s$^{-1}$ cm$^{-2}$,
a typical flux measurement error around the flux limit,
this modification brings a 2\% and 0.5\% increase to $v_{\rm max}(L,T)$
at 1 and 10~keV, respectively,
compared with the simple estimate with Eq. \ref{eq:vmax}.

Although the correction is still insignificant with respect to
the Poisson error
that each temperature bin would have, we use the Eq. \ref{eq:modvmax}
in the derivation of the XTF in Sect. 3.
Our sample selection criteria specify several parameters
in Eq. \ref{eq:diffXTF}, \ref{aeq:VmaxT}, \ref{eq:W} and \ref{eq:modvmax}
as $\Omega$=8.139 str, $f_{\rm obs,lim}$ = $1.99\times 10^{-11}$ ergs s$^{-1}$ cm$^{-2}$
in 0.1-2.4~keV band in the observed frame,
and then $L_{\rm obs}$ should be the 0.1-2.4~keV band luminosity
in the observed frame, which corresponds to a cluster rest frame luminosity
in the energy range from ($1+z_{\rm max})~0.1$ to ($1+z_{\rm max}$)~2.4~keV.
Also, from the $L-T$ relation studied in Sect. 2.3 and Sect. 4,
we estimated the following parameters as
$\log A=42.15$, $\alpha=2.47$, and $\sigma_{\log L}=0.24$
for an open universe case, and
$\log A=42.14$, $\alpha=2.50$, and $\sigma_{\log L}=0.24$
for a flat universe case.
The XTF thus derived is shown in Fig.~\ref{XTF+SIMPLE}.

Finally, we should note that $T_{\rm i}$ in Eq. \ref{eq:diffXTF}
is the measured cluster temperature rather than the ``true'' temperature.
The derived XTF should have been smeared due to
the temperature measurement error,
which brings another systematic error into the resulting XTF.
In order to investigate the possible systematic errors due to
the temperature measurement errors and to confirm our analysis method,
we performed a Monte-Carlo simulation.
With an adopted XTF of $\phi(T) \propto T^{-5}$,
a number of cluster observations are simulated taking into account
the scatter in the $L-T$ relation as well as the flux and temperature
measurement errors, and the XTF was constructed as performed
with the real data using Eqs. \ref{eq:diffXTF}, \ref{aeq:VmaxT},
\ref{eq:W} and \ref{eq:modvmax}.
The input XTF and the rederived XTF from the simulations show complete
agreement except a minor systematic increase --- at most 2\% at 10~keV,
which is due to the temperature measurement error for very hot clusters.
These systematics (Eddington bias; Eddington \cite{Eddington})
remain in the XTF derived in Sect. 3 (Fig.~\ref{XTF+SIMPLE}),
while, for the model fitting in Sect. 4,
the temperature measurement error is taken into account (Eq. 15).

\begin{acknowledgements}
We thank Peter Schuecker, Makoto Hattori, Patrick Henry,
and Paul Lynam for valuable discussions and comments.
We have made use of the {\it ASCA} archival database
at Goddard Space Flight Center/NASA, U.S.A. 
and the Leicester Database and Archive Service
at the Department of Physics and Astronomy,
Leicester University, UK.
We acknowledge FTOOLS.
\end{acknowledgements}

\label{lastpage}

\end{document}